\documentclass[iop]{emulateapj}

\usepackage{epstopdf}
\usepackage{epsfig}
\usepackage{graphics}
\usepackage{graphicx}
\usepackage{natbib}
\newcommand{\Ha}{H$\alpha\ $}
\newcommand{\kms}{km~s$^{-1}$\ }

\begin{document}

\title{Properties of Sequential Chromospheric Brightenings and Associated Flare Ribbons}
\author{Michael S. Kirk\altaffilmark{1,2,3}, K. S. Balasubramaniam\altaffilmark{2,3,1}, Jason Jackiewicz\altaffilmark{1}, R. T. James McAteer\altaffilmark{1}, Ryan O. Milligan\altaffilmark{4}}
\altaffiltext{1}{Department of Astronomy, New Mexico State University, P.O. Box 30001, MSC 4500, Las Cruces, New Mexico 88003-8001; mskirk@nmsu.edu}
\altaffiltext{2}{Space Vehicles Directorate, Air Force Research Laboratory, Kirtland AFB, NM 87114}
\altaffiltext{3}{National Solar Observatory, Sunspot, NM 88349}
\altaffiltext{4}{Queen's University Belfast, University Road Belfast, BT7 1NN, Northern Ireland, UK}

\begin{abstract}
We report on the physical properties of solar sequential chromospheric brightenings (SCBs) observed in conjunction with moderate-sized chromospheric flares with associated CMEs. To characterize these ephemeral events, we developed automated procedures to identify and track subsections (kernels) of solar flares and associated SCBs using high resolution \Ha images. Following the algorithmic identification and a statistical analysis, we compare and find the following: SCBs are distinctly different from flare kernels in their temporal characteristics of intensity, Doppler structure, duration, and location properties. We demonstrate that flare ribbons are themselves made up of subsections exhibiting differing characteristics. Flare kernels are measured to have a mean propagation speed of 0.2~\kms and a maximum speed of 2.3~\kms over a mean distance of $5\times10^3$~km. Within the studied population of SCBs, different classes of characteristics are observed with coincident negative, positive, or both negative and positive Doppler shifts of a few km~s$^{-1}$.  The appearance of SCBs precede peak flare intensity by $\approx 12$ minutes and decay $\approx 1$ hour later.  They are also found to propagate laterally away from flare center in clusters at 41~\kms or 89~km~s$^{-1}$. Given SCBs distinctive nature compared to flares, we suggest a different physical mechanism relating to their origin than the associated flare. We present a heuristic model of the origin of SCBs. 
\end{abstract}


\section{Introduction}
The solar chromosphere exhibits three different classes of small scale intensity brightenings: flare-, plage-, and compact-brightenings. Although each is characterized by an enhanced temporal \Ha brightness relative to a background quiet Sun, they each have distinct physical processes governing their spatial and temporal evolution. Typically brightenings have been identified and characterized manually from a single data source~\citep[e.g.,][]{Kurt2000,Ruzdjak1989,Veronig2002}.  However in order to form a better understanding of the underlying dynamics,  data from multiple sources must be utilized and numerous similar features must be statistically analyzed. 

This work focuses on flare brightenings and associated compact brightenings called sequential chromospheric brightenings (SCBs). SCBs were first observed in 2005 and appear as a series of spatially separated points that brighten in sequence \citep{Bala2005}. SCBs are observed as multiple trains of brightenings in association with a large-scale eruption in the chromosphere or corona and are interpreted as progressive propagating disturbances. The loci of brightenings emerge predominantly along the axis of the flare ribbons. SCBs are correlated with the dynamics which cause solar flares, coronal restructuring of magnetic fields, halo CMEs, EIT waves, and chromospheric sympathetic flaring \citep{Bala2005}. \citet{Pevtsov2007} demonstrate that SCBs have properties consistent with aspects of chromospheric evaporation.

This article presents a new description of the dynamical properties of SCBs resulting from applying a new automated method~\citep{Kirk2011} of identifying and tracking SCBs and associated flare ribbons. This tracking technique differs from previous flare tracking algorithms in that it identifies and tracks spatial and temporal subsections of the flare and all related brightenings from pre-flare stage, through the impulsive brightening stage, and into their decay. Such an automated measurement allows for tracking dynamical changes in intensity, position, and derived Doppler velocities of each individual subsection. The tracking algorithm is also adapted to follow the temporal evolution of ephemeral SCBs that appear with the flare. In Section~\ref{S-Data} we describe the data used to train the algorithm and the image processing involved in the detection routine. In Section~\ref{S-Flare} we present the results of tracking the evolution of flare kernels through an erupting flare. In Section~\ref{S-SCB} we present the application of the tracking algorithm to ephemeral SCBs.  We present a physical interpretation of the SCBs and provide a heuristic a model of the origin of SCBs in Section~\ref{S-Interp}.  Finally, in Section~\ref{S-Discussion} we discuss implications of these results and provide future direction for this work.

\section{Data and data processing}
\label{S-Data}

\begin{figure}    
   \centerline{\includegraphics[width=0.8\textwidth,clip=,angle=0]{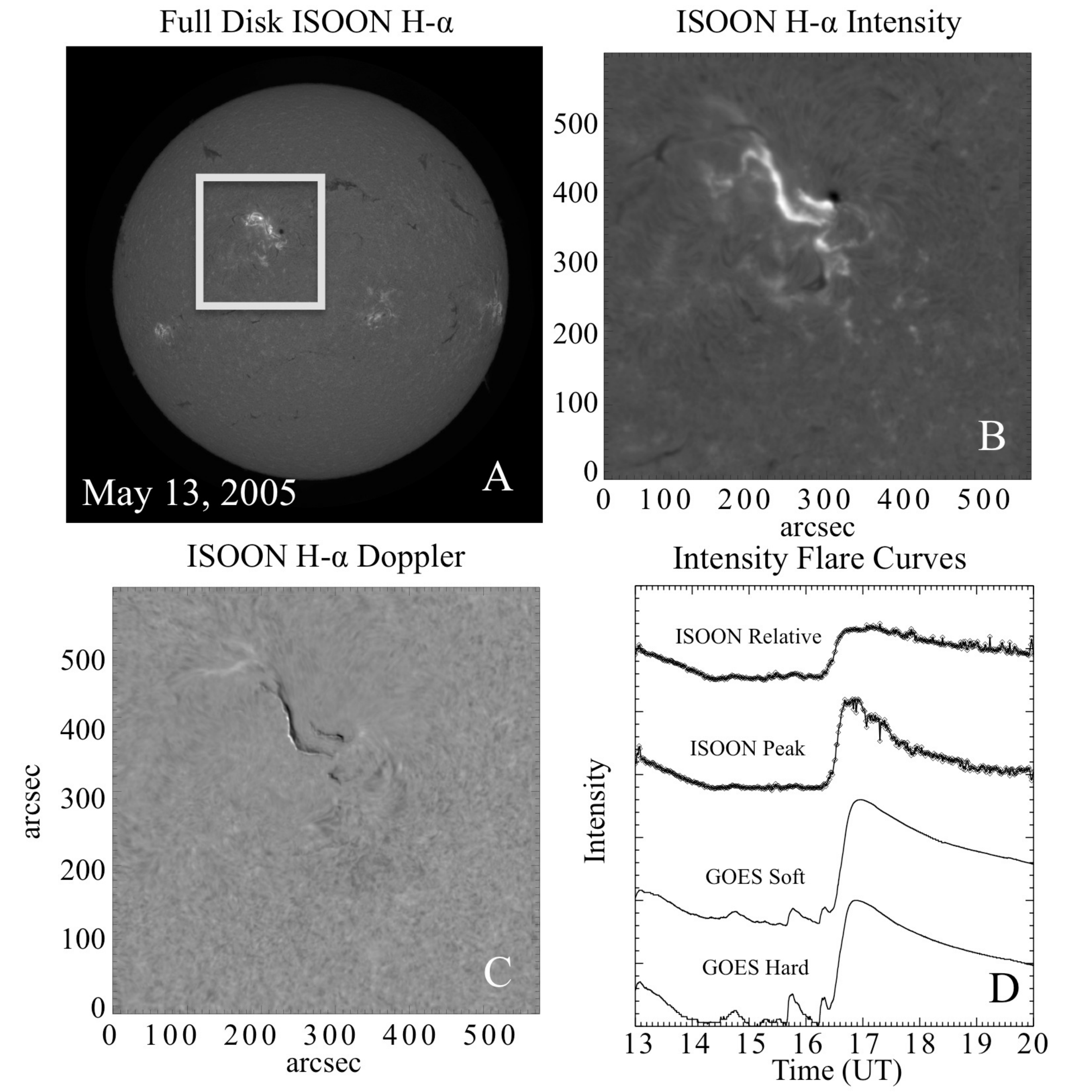}} 

              \caption{The two-ribbon flaring event from 13 May 2005. {\bf A} shows an example of a calibrated H$\alpha$ ISOON image with the region of interest (ROI) highlighted. The ISOON images are 2048 $\times$ 2048 pixels with each pixel having a 1.1 arcsec resolution. {\bf B} is the ROI after preprocessing. This ROI coves 576 $\times$ 576 pixels corresponding to $2.12 \times 10^{11}$ km$^2$ on the solar surface.  {\bf C} is a Doppler measure of the same ROI as part {\bf B}. The Doppler velocity image is created out of ISOON H$\alpha$ off-band images described in Section~\ref{S-Data} where the velocity ranges from -26.6 to 21.5 \kms from black to white.  {\bf D} shows intensity curves over the time period of interest: 13:00--20:02 UT as described in detail in the text.}
   \label{0513_boxes}
   \end{figure}

\begin{figure}    
     \centerline{\includegraphics[width=0.8\textwidth,clip=,angle=0]{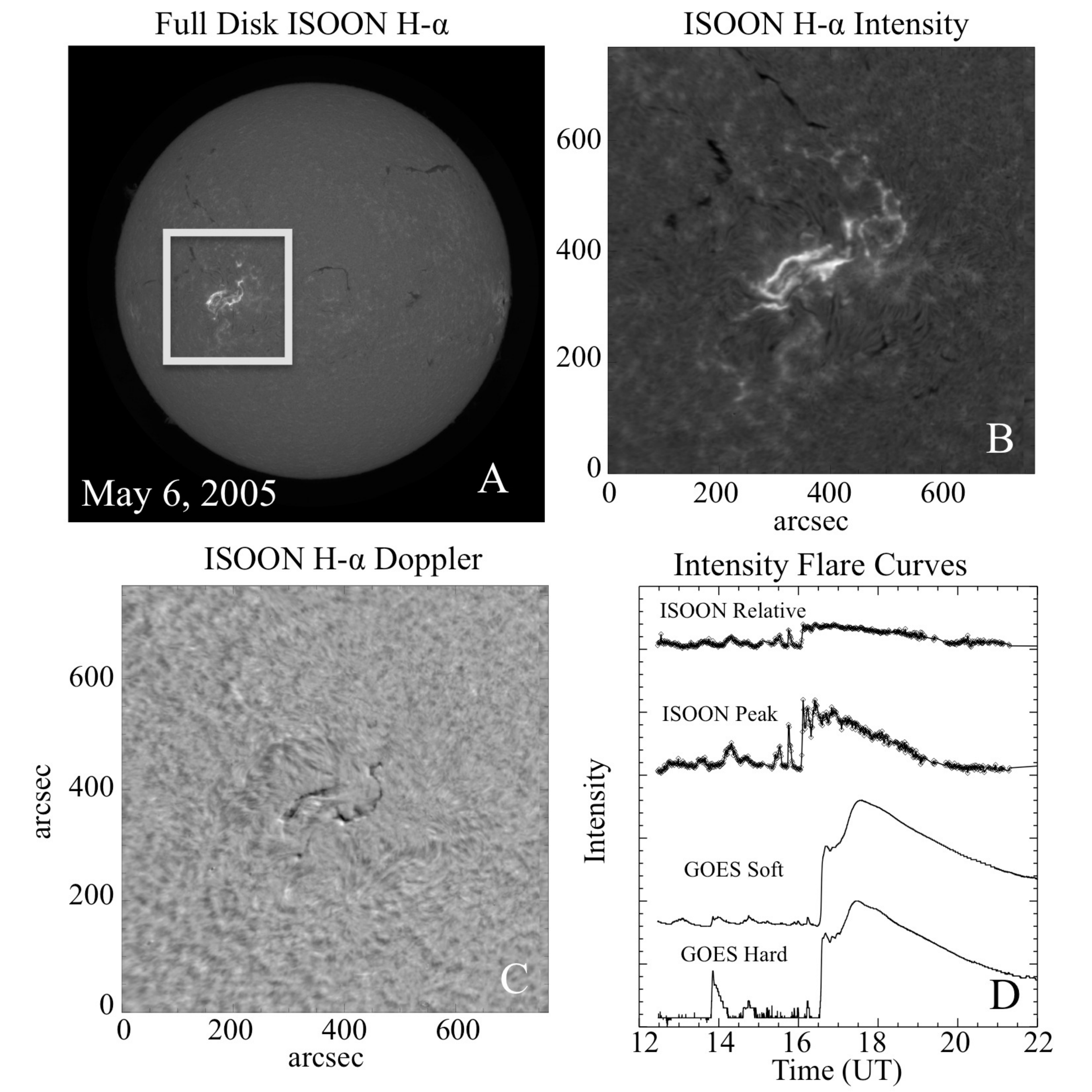}
              }
              \caption{The two-ribbon flaring event from 6 May 2005. {\bf A} shows an example of a calibrated H$\alpha$ ISOON image with the region of interest (ROI) highlighted. The ISOON images are 2048 $\times$ 2048 pixels with each pixel having a 1.1 arcsec resolution. {\bf B} is the ROI after preprocessing. This ROI coves 768 $\times$ 768 pixels corresponding to $3.77 \times 10^{11}$ km$^2$ on the solar surface.  {\bf C} is a Doppler measure of the same ROI as part {\bf B}. The Doppler velocity image is created out of ISOON H$\alpha$ off-band images described in Section~\ref{S-Data} where the velocity ranges from -12.5 to 6.9 \kms from black to white.  {\bf D} shows intensity curves over the time period of interest: 12:28--22:47 UT as described in detail in the text.}
   \label{0506_boxes}
   \end{figure}

\begin{figure}    
	\centerline{\includegraphics[width=0.8\textwidth,clip=,angle=0]{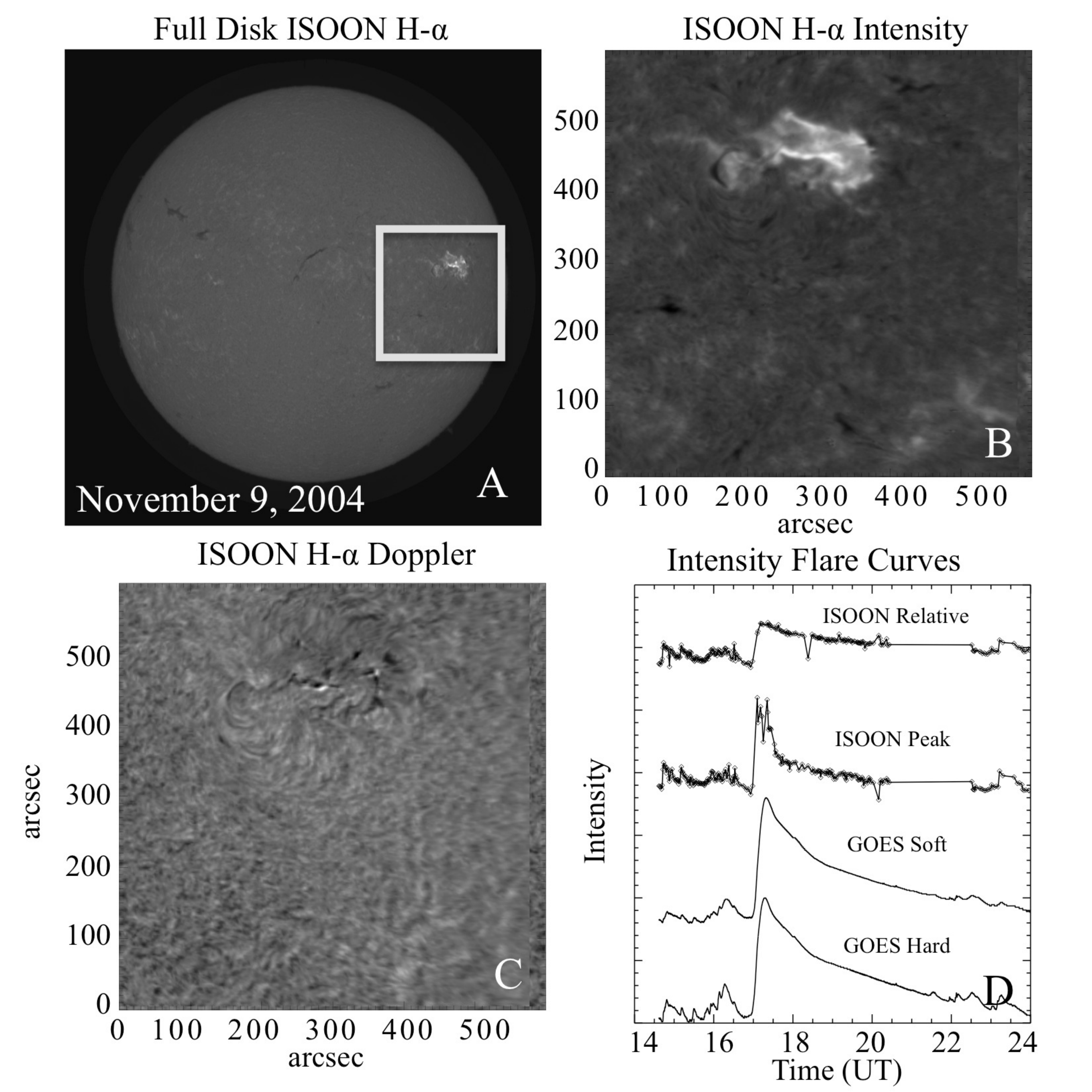}
              }
              \caption{The two-ribbon flaring event from 9 November, 2004. {\bf A} shows an example of a calibrated H$\alpha$ ISOON image with the region of interest (ROI) highlighted. The ISOON images are 2048 $\times$ 2048 pixels with each pixel having a 1.1 arcsec resolution. {\bf B} is the ROI after preprocessing. This ROI coves 600 $\times$ 600 pixels corresponding to $2.30 \times 10^{11}$ km$^2$ on the solar surface.  {\bf C} is a Doppler measure of the same ROI as part {\bf B}. The Doppler velocity image is created out of ISOON H$\alpha$ off-band images described in Section~\ref{S-Data} where the velocity ranges from -9.3 to 14.6 \kms from black to white.  {\bf D} shows intensity curves over the time period of interest: 14:37 -- 23:58 UT as described in detail in the text. }
   \label{1109_boxes}
   \end{figure}

In this study we use chromospheric \Ha (6562.8 \AA) images from the USAF's {\it Improved Solar Observing Optical Network} (ISOON;~\citealp{Neidig1998}) prototype telescope to study flare ribbons and SCBs. ISOON is an automated telescope producing 2048$\times$2048 pixel full-disk images at a one-minute cadence. Each image has a 1.1 arc-second spatial sampling, is normalized to the quiet Sun, and corrected for atmospheric refraction (Figures~\ref{0513_boxes}(a), ~\ref{0506_boxes}(a), and \ref{1109_boxes}(a)). Nearly coincident to the spectral line center images, ISOON also records \Ha $\pm0.4$ \AA\ off-band Doppler images, from which Doppler signals are derived. 

For this study, we chose to apply the brightening detection algorithms to three flares where~\citet{Bala2006} had previously identified SCB events (Table~\ref{T-events}). Each of the events selected has a two-ribbon configuration and an associated halo CME. Of these, both the 6 May 2005 and 13 May 2005 events were both located near disk center while the 9 November 2004 event was near the western limb. Images were extracted from the archive from $\pm 3.5$ hours of the eruption start time, yielding a data cube with $\approx 400$ images for each event.  An \Ha relative intensity curve, an \Ha maximum intensity curve, and {\it Geostationary Operational Environmental Satellites} (GOES) hard and soft x-ray fluxes are plotted to characterize the flare (Figures~\ref{0513_boxes}(d), ~\ref{0506_boxes}(d), and \ref{1109_boxes}(d)) as described in~\citet{Kirk2011}. 

\begin{table}
\caption{ 
The events selected for study. Each event was visually identified as having a two-ribbon configuration and SCBs associated with the flare. 
}
\label{T-events}

\begin{tabular}{lcccc}     
  \hline                   
 Date & Start (UT) & Duration (h) & Flare Class & CME \\
  \hline
9 Nov. 2004& 16:59 & 0.5 & M8.9 & Halo \\ 
6 May 2005& 16:03 & 2.1 & C8.5& Halo \\
13 May 2005& 16:13 & 1.3 & M8.0& Halo \\

  \hline
\end{tabular}
\end{table}

Each ISOON image is reconditioned to remove solar limb-darkening. The images are then de-projected into conformal coordinates using a Guyou projection~\citep[an oblique aspect of the Peirce projection,][]{Peirce1879}, which removes the projection effects of imaging the solar sphere.   Each image is then cropped to the region of interest (ROI) (Figures~\ref{0513_boxes}(b), ~\ref{0506_boxes}(b), and \ref{1109_boxes}(b)).  The set of images are aligned using a cross-correlation algorithm eliminating the rotation effects of the Sun. Frames that contain bad pixels or excessive cloud interference are removed. 

The red and blue wings of ISOON Doppler images are each preprocessed using the same technique as applied to the line center images described above (Figures~\ref{0513_boxes}(c), ~\ref{0506_boxes}(c), and \ref{1109_boxes}(c)). In order to produce velocity measurements, a Doppler cancelation technique is employed in which red images are subtracted from blue images \citep[for a modern example see:][]{Connes1985}. Typically, red and blue images are separated by about 4 seconds and no more than 5 seconds. A mean zero redshift in the subtracted Doppler image in the quiet-Sun serves as a reference. In dynamical situations such as solar flares, \Ha profiles are often asymmetric especially in emission making this Doppler technique invalid. To avoid asymmetric profiles, only values derived outside of the flaring region are considered such that the areas of interest at $\pm 0.4$ \AA\ are still in absorption even if raised in intensity 
\citep{Bala2004}. The values from the subtracted images are translated into units of \kms by comparison to a measured response in spectral line shifts, which are calibrated against an intensity difference for the ISOON telescope. In this context, we interpret the Doppler shift as a line of sight velocity. Thus, Doppler velocity [$v_D$] is defined as:
\begin{equation}
v_D=k\left(\mathcal{I}_{\rm blue} - \mathcal{I}_{\rm red}\right),
\end{equation}
where $\mathcal{I}$ is the measured intensity and $k$ is a linear fitted factor that assumes the intensity changes due to shifts in the symmetric spectral line, which can be attributed to a Doppler measure as a first-order approximation. The linear factor [$k$] has been independently determined for the ISOON telescope by measuring the full \Ha spectral line Doppler shift across the entire solar disk~\citep{Bala2011}.

\subsection{Detection and tracking}
\label{S-detection}

An animated time series of sequential images covering an erupting flare reveals several physical characteristics of evolving ribbons: the ribbons separate, brighten, and change their morphology. Adjacent to the eruption, SCBs can be observed brightening and dimming in the vicinity of the ribbons.  \citet{Kirk2011} describe in detail techniques and methods used to extract quantities of interest such as location, velocity, and intensity of flare ribbons and SCBs. The thresholding, image enhancement, and feature identification are tuned to the ISOON data. The detection and tracking algorithms are specialized for each feature of interest and requires physical knowledge (e.g. size, peak intensity, and longevity) of that feature being detected to isolate the substructure. 

Briefly, the detection algorithm first identifies candidate bright kernels in a set of images. In this context, we define a kernel to be a locus of pixels that are associated with each other through increased intensity as compared with the immediately surrounding pixels. Each kernel has a local maximum, must be separated from another kernel by at least one pixel, and does not have any predetermined size or shape. Next, the algorithm links time-resolved kernels between frames into trajectories.  A filter is applied to eliminate inconsistent or otherwise peculiar detections. Finally, characteristics of bright kernels are extracted by overlaying the trajectories over complementary datasets. To aid in this detection, tracking software developed by Crocker, Grier and Weeks was used as a foundation and modified to fit the needs of this project\footnote[1]{Crocker's software is available online at \url{www.physics.emory.edu/$\sim$weeks/idl/}.} \citep{Crocker1996}.  

\begin{figure}    
   \centerline{\includegraphics[width=1.0\textwidth,clip=]{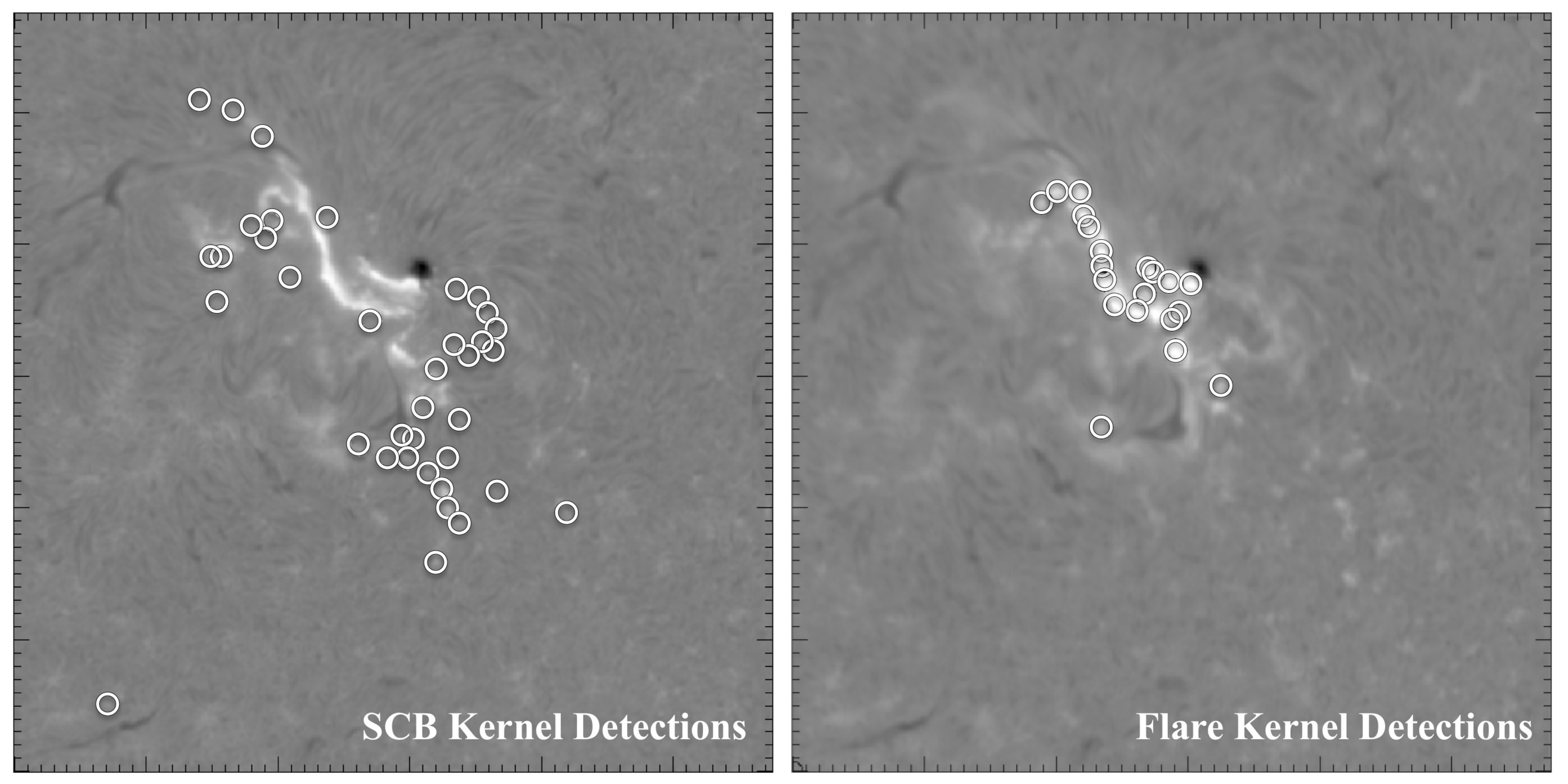}
              }
              \caption{Example ISOON H$\alpha$ images from the 13 May 2005 two ribbon flare. The detection algorithm is applied to the image series to extract either ribbon brightening or compact brightening. The left image shows one time slice of individual SCBs identified external to the flare ribbon. The right image shows the same time slice with brightenings identified along the flare ribbon. Each detection is made up of a small matrix of pixels associated with the brightening called a kernel. 
               }
   \label{Kernel_marks}
   \end{figure}

\begin{figure}    
   \centerline{\includegraphics[width=0.5\textwidth,clip=]{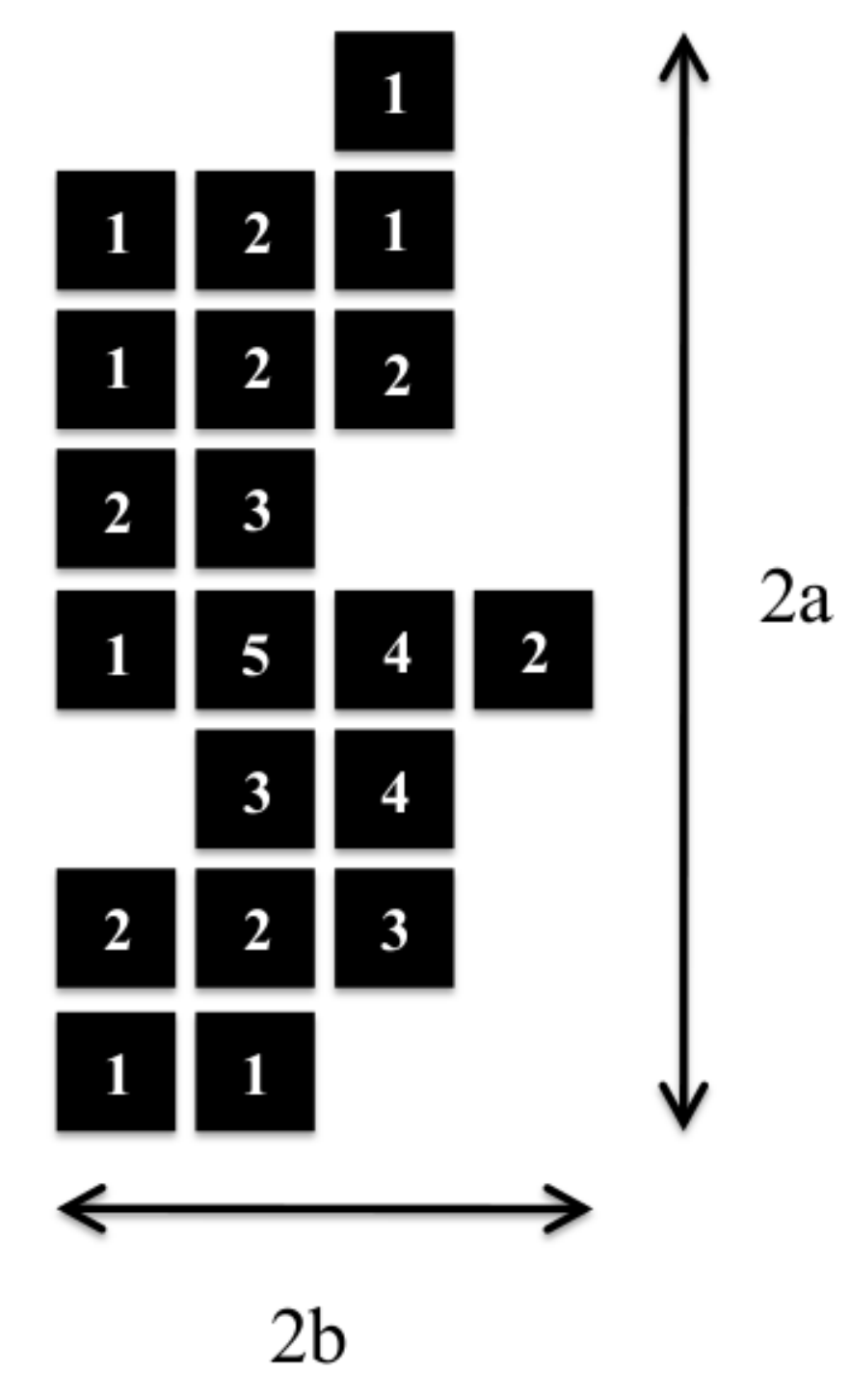}
              }
              \caption{A diagram of a hypothetical kernel (either flare or SCB) used for classifications in this paper. Each pixel is labeled with a representative intensity [$\mathcal{A}_i$]. The semi-major axis [$a$] and semi-minor axis [$b$] are shown. This kernel has the following properties: total intensity, $m_0=43$, semi-major axis, $a=4$, semi-minor axis, $b=2$, eccentricity, $e=0.87$, centroid, $(x,y)=(0.26,0.18)$, and radius of gyration, $R_g=1.9$. The maximum intensity, in this case 5, is defined as having pixel coordinates $(0,0).$
                                          }
   \label{Kernel_diagram}
   \end{figure}

In order to characterize a kernel, we calculate its integrated intensity, radius of gyration, and eccentricity. The eccentricity of the kernel (as defined by its semi-major [a] and semi-minor [b] axis) is calculated using, 
\begin{equation}
\label{Eq-ecc}
e=\sqrt{1-\frac{b^2}{a^2}}.
\end{equation}
The integrated intensity of a given kernel is defined by,
\begin{equation}
\label{Eq-tot_inten}
m= \sum_{i^2+j^2 \le \omega^2}\mathcal{A}(x+i, y+j)
\end{equation}
where $\mathcal{A}(x+i,y+j)$ is the intensity of the pixel located at $(x+i,y+j)$, the kernel has a brightness weighted centroid with coordinates $(x,y)$, and $\omega$ is the radius of the mask \citep{Crocker1996}. The radius of the mask is chosen to be eight pixels in this case. A radius of gyration, $R_g$, is related to the moment of inertia [$I$] by using:
\begin{equation}
\label{Eq-inertia}
I = \sum_{k} m_k  r_k^2 = m  R_g^2, 
\end{equation}
\begin{equation}
r=(a+b)/2,
\end{equation}
where $m_k$ is the mass of particle $k$, $m$ is the total mass of the system, and $r$ is the distance to the rotation axis. In this context, we interpret the radius of gyration to be
\begin{equation}
\label{Eq-radius}
R_g^2 = \sum_{k} \frac{\mathcal{A}_k r_k^2}{m} 
\end{equation}
where $m$ is the integrated intensity as defined in Equation~\ref{Eq-tot_inten} \citep{Crocker1996}. Figure~\ref{Kernel_diagram} diagrams a hypothetical kernel.  The results of this algorithm identify and characterize several kernels in the flaring region (Figure~\ref{Kernel_marks}). The total number of flare kernels detected is typically between 100 -- 200, while the number of SCB kernels detected is typically three to four times that number. 

\section{Flare ribbon properties}
\label{S-Flare}
The properties of kernels identified in flare ribbons can be examined in two ways: (i) each kernel can be considered as an independent aggregation of compact brightenings or (ii) kernels can be considered as dependent on each other as fragments of a dynamic system. Each category brings about contrasting properties. If one considers the kernels as independent elements, this provides a way to examine changes to subsections of the flare and is discussed in Section~\ref{S-Individual}. Associating kernels with their contextual surroundings allows a way to examine the total evolution of the flare without concern of how individual kernels behave. This type of examination is addressed in Section~\ref{S-total}. 

\subsection{Qualities of individual kernels}
\label{S-Individual}

\begin{figure}    
   \centerline{\includegraphics[width=0.7\textwidth,clip=]{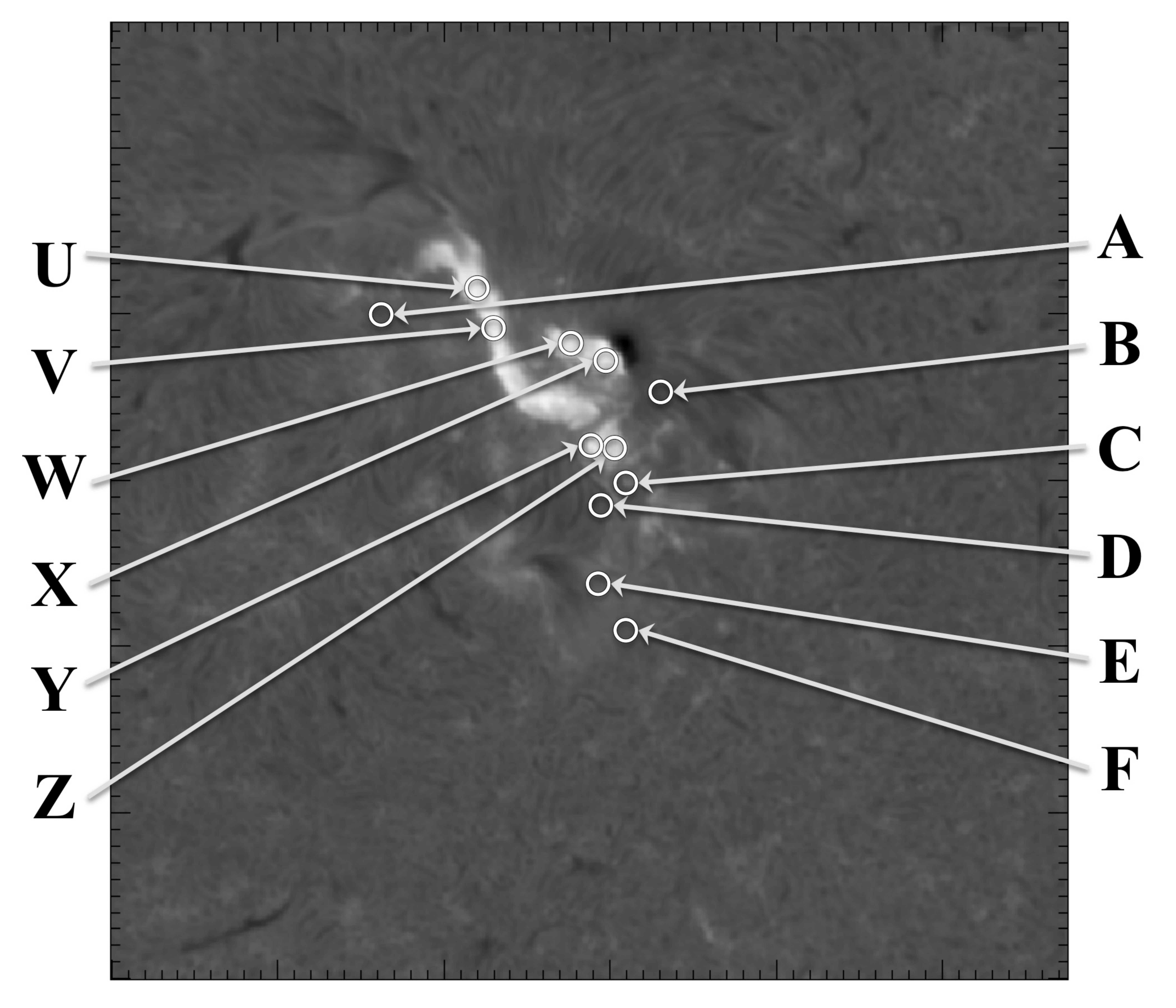}
              }
              \caption{An image from the 13 May 2005 event. Letters A -- F point to the locations several SCB kernels that are shown in more detail in Figure~\ref{Kernel_curves}. Letters U -- Z point to the locations of a sample of several flare kernels that are also shown in more detail in Figure~\ref{Kernel_curves}. Note that since flare kernels move with the evolving flare ribbon, the locations of these kernels are only accurate for this image. In contrast, SCB kernels are stationary and remain in the same location in any image. 
                             }
   \label{Kernel_map}
   \end{figure}

A sample of six kernels from the 13 May 2005 event, letters U - X in Figure~\ref{Kernel_map}, is representative of the majority of flare kernels tracked. The kernels were selected from three different regions of this flare. The normalized \Ha intensity of each of these kernels is shown in Figure~\ref{Kernel_curves}.  In contrast with the integrated flare light curve, an individual flare kernel often has a shorter lifetime. This is because an individual kernel may first appear in the impulsive phase of the flare (as demonstrated in kernels W - Y) while other kernels may disappear as the flare begins to decay (as in kernels U and Z). Most likely these kernels merged with one another, at which point their unique identity was lost. Kernels W and X are the only two that show the characteristic exponential dimming found in the reference curves in Figure~\ref{0513_boxes}~(d). All of these kernels have peak intensities within a few minutes of the peak of the total flare intensity and have sustained brightening above background levels for up to an hour. 

The integrated speed of displaced kernels provides context to the evolution of intensities. We define the integrated speed of displacement to be the sum over all time steps of the measured velocity of an individual kernel and thus small velocity perturbations are minimized by the sum. The peak integrated speed measured for each kernel peaks at $\sim 2.3$ km s$^{-1}$ and has a mean of $\sim 0.2$ km s$^{-1}$. There is significant motion along the flare ribbons as well as outflow away from flare center. The motions are complex but generally diverge. The total spatial (angular) displacement of each flare kernel peaks at $\sim 20.7$ Mm and has a mean of $\sim 5.0 $ Mm. Flare kernels that exhibited the greatest speeds did not necessarily have the greatest displacements. As the flare develops, most of the motion in flare ribbons indicated is synchronous with the separation of the ribbons. A few tracked kernels indicate motion along flare ribbons. These flows along flare ribbons are consistently observed in all three flares. 

\subsection{Derived flare quantities}
\label{S-total}

\begin{figure}    
   \centerline{\includegraphics[width=1.0\textwidth,clip=,angle=0]{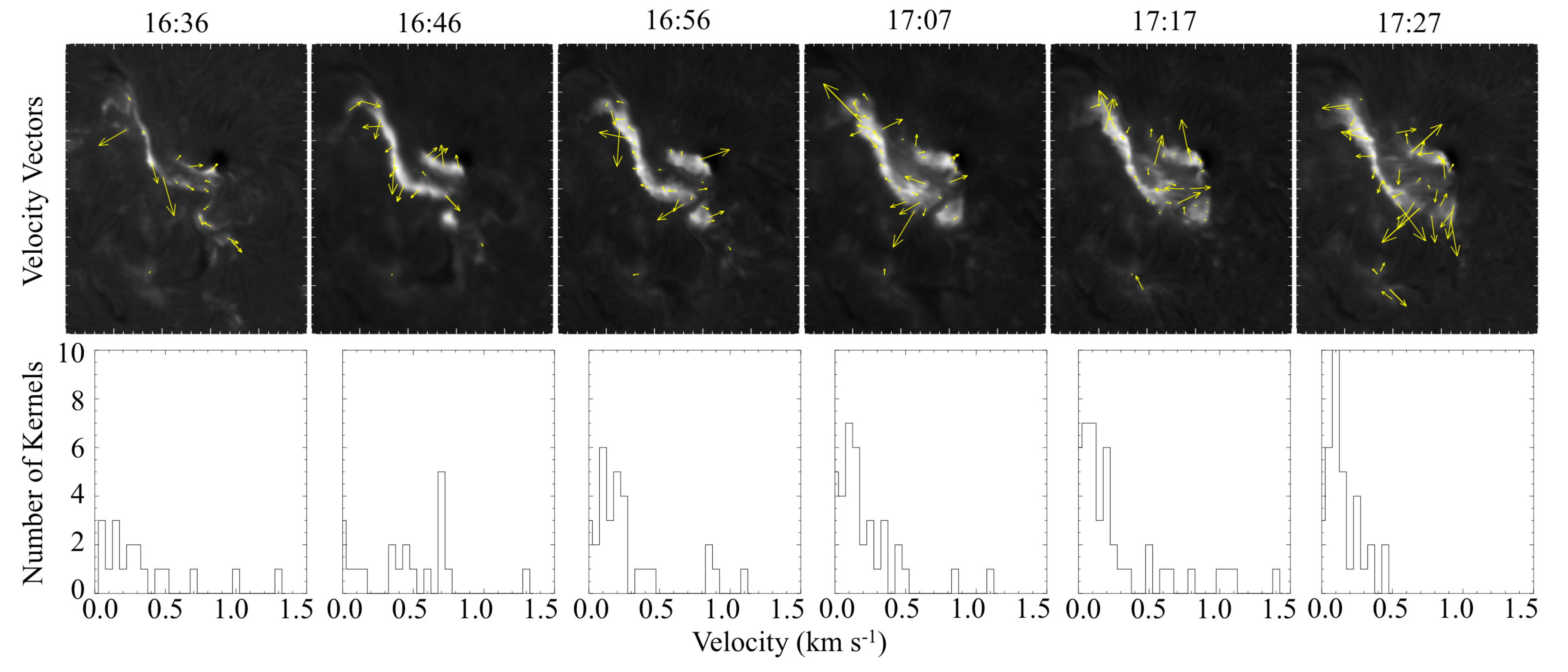}
              }
              \caption{A time series of images with the apparent lateral velocity of each tracked flare kernel at that time step plotted over the H$\alpha$ flare ribbons. A histogram of the number of flare kernels as a function of velocity corresponding to each time step is plotted underneath.  
                             }
   \label{Flare_vel}
   \end{figure}

Examining trajectories of flare kernels offers insight into the motions of the flare ribbons as they evolve through the eruption. Figure~\ref{Flare_vel} shows a time series of images centered about the peak-time of the flare. The velocities of the detected flare kernels are superposed as vectors. Notice the initial outflow of the two ribbons near the flare peak at 16:49 UT. But, as the flare continues to evolve, there is more motion along the flare ribbons. This is probably a result of the over-arching loops readjusting after the reconnection event.  Beneath each image is a histogram of the distribution of velocities at each time step. The mean apparent lateral velocity of each kernel remains below 0.5 \kms throughout the flare even though the total velocity changes significantly. Qualitatively, the bulk of the apparent motion appears after the flare peak intensity at 16:49 UT. 

\begin{figure}    
   \centerline{\includegraphics[width=0.7\textwidth,clip=,angle=0]{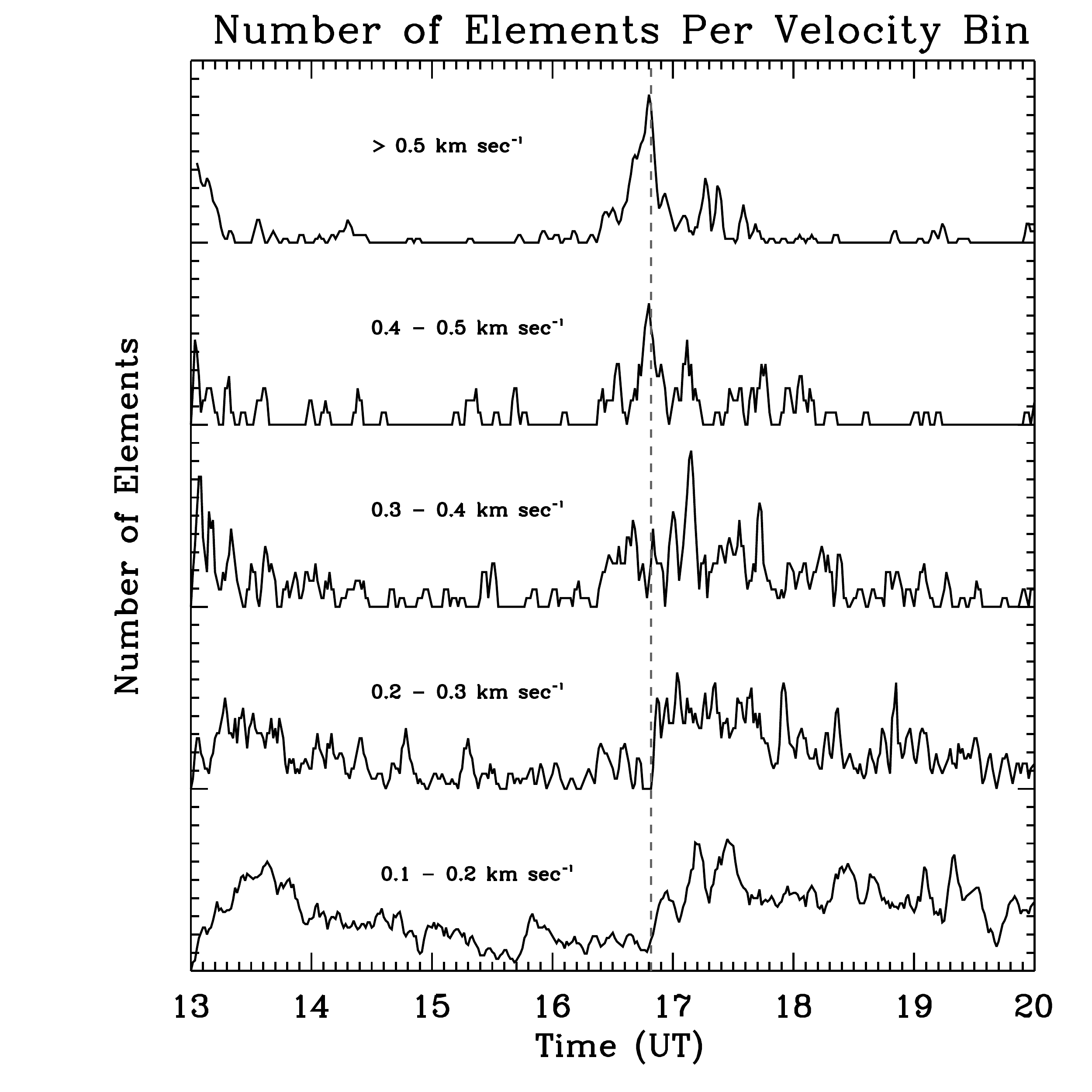}
              }
              \caption{The number of elements in each apparent lateral velocity bin plotted as a function of time. The vertical dashed line marks the time of the peak flare intensity. The velocity bins are 0.1 km s$^{-1}$ in width. The smallest bin, 0--0.1 km s$^{-1}$, is excluded because of a base level of noise inherent in the detection routine.   
                             }
   \label{Flare_vel_bins}
   \end{figure}

Grouping the kernels' velocities into appropriate bins provides another way to analyze the dynamics of the flare (Figure~\ref{Flare_vel_bins}). The kernels with apparent lateral speeds above 0.4 \kms are highly coincident with the peak intensity of the flare. The number of these fast moving kernels peaks within a couple minutes of flare-peak, and then quickly decays back to quiet levels. The velocity bins between 0.1 and 0.4 \kms show a different substructure. These velocity bins peak $\approx 30$ minutes after the peak of the flare intensity. They show a much slower decay rate, staying above the pre-flare velocity measured throughout the decay phase of the \Ha flare curve. 

\begin{figure}    
   \centerline{\includegraphics[width=1.0\textwidth,clip=]{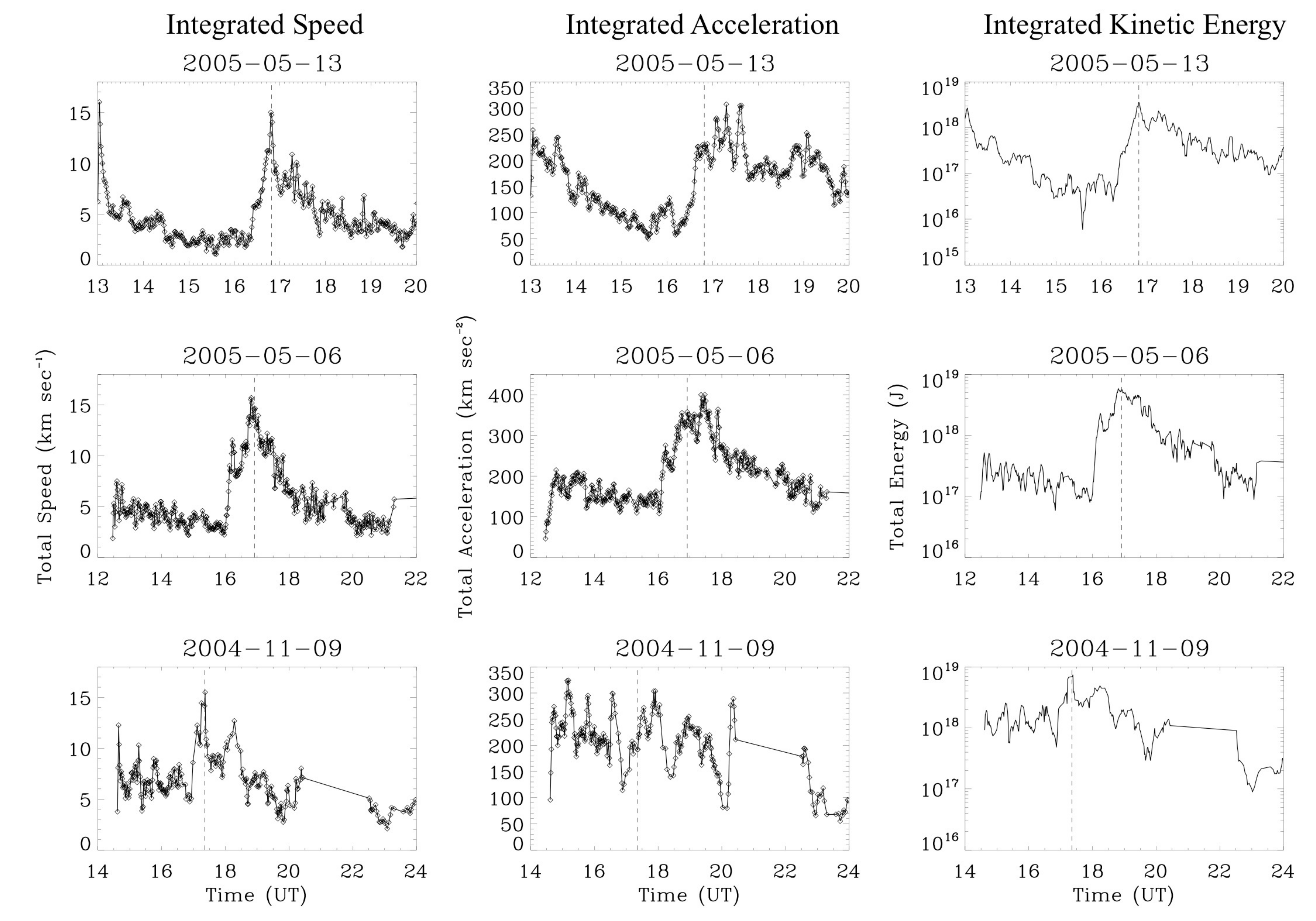}
                 }
              \caption{ The integrated unsigned propagation speed, acceleration, and energy of flare kernels as a function of time for each of the three events studied.  The vertical dashed line indicates the maximum flare intensity for the flaring event.  The first column shows the integrated speed curves. The second column shows the acceleration of the flare ribbons. The third column shows the integrated kinetic energy, equation~\ref{eq-energy}, as a function of time for each of the three events studied. The 2004 flare had significant noise in the original data and a location on the limb contributing to the inconsistent measurements. 
                            }
   \label{Integrated_speed}
   \end{figure}

Figure~\ref{Integrated_speed} (left column) shows the integrated unsigned kernel velocity (speed) for each image. In general, the evolution of the integrated kernel speed has a shape similar to the intensity curve. For the 6 May 2005 event, the speed curve is more similar to the GOES x-ray intensity curve than the \Ha curve. For the November 2004 event, the limb geometry of the flare, as well as the noise in the original dataset, results in a noisy curve. Despite these difficulties, a clear increase in total speed is apparent near the peak of the flare. A continuum integrated speed level of 1--2 \kms is consistent in each event, indicating the quality of data and of the tracking software. The peak integrated speed is about 15 \kms for all three flares. 

Examining the change in the speed of each kernel between successive images yields a derived acceleration for each tracked kernel. The integrated unsigned acceleration is plotted in Figure~\ref{Integrated_speed} (middle column). Both the 13 May 2005 and 6 May 2005 events show peak acceleration after the peak of the intensity curve. Hence the majority of the acceleration in the apparent motion of the flare comes in the decay phase of \Ha emission, in concert with the formation of post-peak flare loops, as seen in movies of TRACE images of other flares. The peak acceleration appears uncorrelated with the strength of the flare since the peak of the M class 13 May event is just over 300 km s$^{-2}$, while the peak of the C class flare on 6 May is nearly 400 km s$^{-2}$. The acceleration curve for the 9 November flare has too much noise associated with it to decisively determine the peak value. 

The right column of Figure~\ref{Integrated_speed} shows a derived kinetic energy associated with the measured motion of the flare kernels. This was accomplished by assuming a chromospheric density [$\sigma$] of $10^{-5}$ kg m$^{-3}$ and a depth [h] of 1000 km. The mass contained in each volume is then calculated by
\begin{equation}
m=\pi h \sigma R_g^2 \sqrt{1-e^2} 
\end{equation}
where $R_g$ is defined in Equation~\ref{Eq-radius} and $e$ is defined in Equation~\ref{Eq-ecc}.  The kinetic energy under a kernel is defined in the standard way to be
\begin{equation}
\label{eq-energy}
E_k=\frac{1}{2} m v^2
\end{equation}
where $v$ (speed) is a measured quantity for each kernel. The derived kinetic energy curves show a $\approx$ 30 fold increase in kinetic energy during the flare with a decay time similar to the decay rate of the \Ha intensity. The measure of a kernel's kinetic energy is imprecise because the measured motion is apparent motion of the underlying plasma. In the model of two ribbon solar flares put forth by~\citet{Priest2002}, the apparent velocity is a better indicator of rates of magnetic reconnection rather than of plasma motion. Despite this caveat, the derived kernel kinetic energy is a useful measure because it combines the size of the kernels with their apparent velocities to characterize the flare's evolution, thus representing the two in a quantitative way.  

\begin{figure}    
   \centerline{\includegraphics[width=0.80\textwidth,clip=,angle=0]{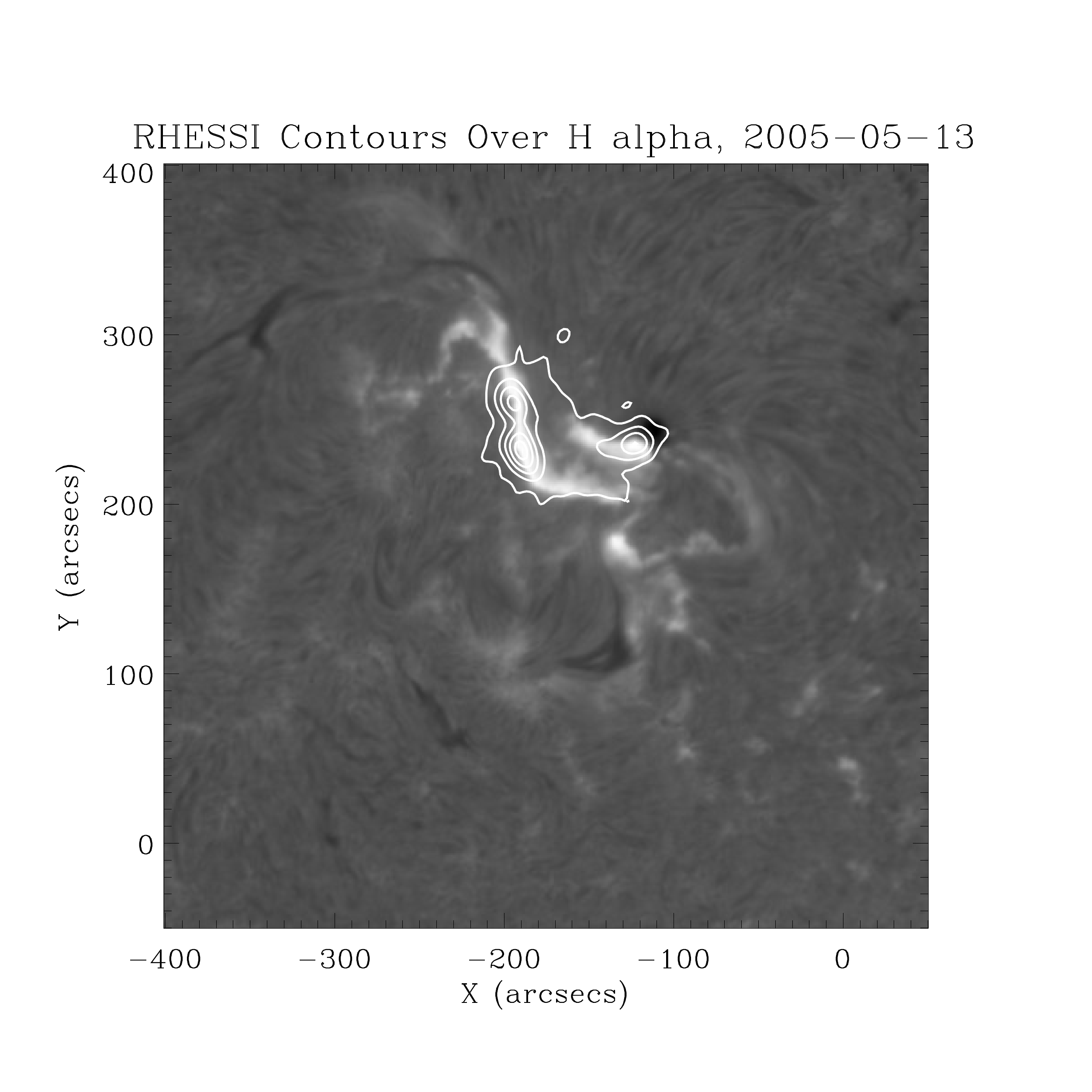}
              }
              \caption{An H$\alpha$ image of the May 13, 2005 flare at peak intensity is shown with the reconstructed RHESSI 25 -- 50 keV pass band integrated between 16:42--16:43 UT contoured on top at 10, 30, 50, 70, and 90\% of the maximum intensity. The field of view of the RHESSI image [-220:-95, 180:305] is confined within the flaring region. }
   \label{RHESSI_over}
   \end{figure}

The non-thermal 25--50 keV emission between 16:42--16:43 UT from the {\it Reuven Ramaty High Energy Solar Spectroscopic Imager} (RHESSI,~\citealp{Lin2002}) of the 13 May flare is contoured in Figure~\ref{RHESSI_over} over an \Ha image taken at 16:42 UT. The high-energy emission is centered over the flare ribbons but is discontinuous across the ribbon. There are three localized points from which the majority of the x-ray emission comes. The parts of the flare ribbons exhibiting the most displacement show generally lower x-ray intensity than the more stationary segments of the flare ribbons. 

\section{SCB properties}
\label{S-SCB}

The measured properties of SCBs can be studied using two scenarios: each SCB kernel is considered as independent and isolated compact brightenings; each compact brightening is considered to be fragments of a dynamic system that has larger structure. This is similar to the approach used for flares.  Again, considering both scenarios has its benefits.  Identifying kernels as independent elements reveals structural differences in compact brightenings and is discussed in Section~\ref{S-IndividualSCB}. Associating kernels with their contextual surroundings provides some insight into the physical structure causing SCBs. An aggregate assessment  of SCBs is addressed in Section~\ref{S-totalSCB}. To mitigate the effects of false or marginal detections, only SCBs with intensities two standard deviations above the mean background intensity are considered for characterizing the nature of the ephemeral brightening.

\subsection{Qualities of individual SCBs}
\label{S-IndividualSCB}

\begin{figure}    
   \centerline{\includegraphics[width=1.0\textwidth,clip=,angle=0]{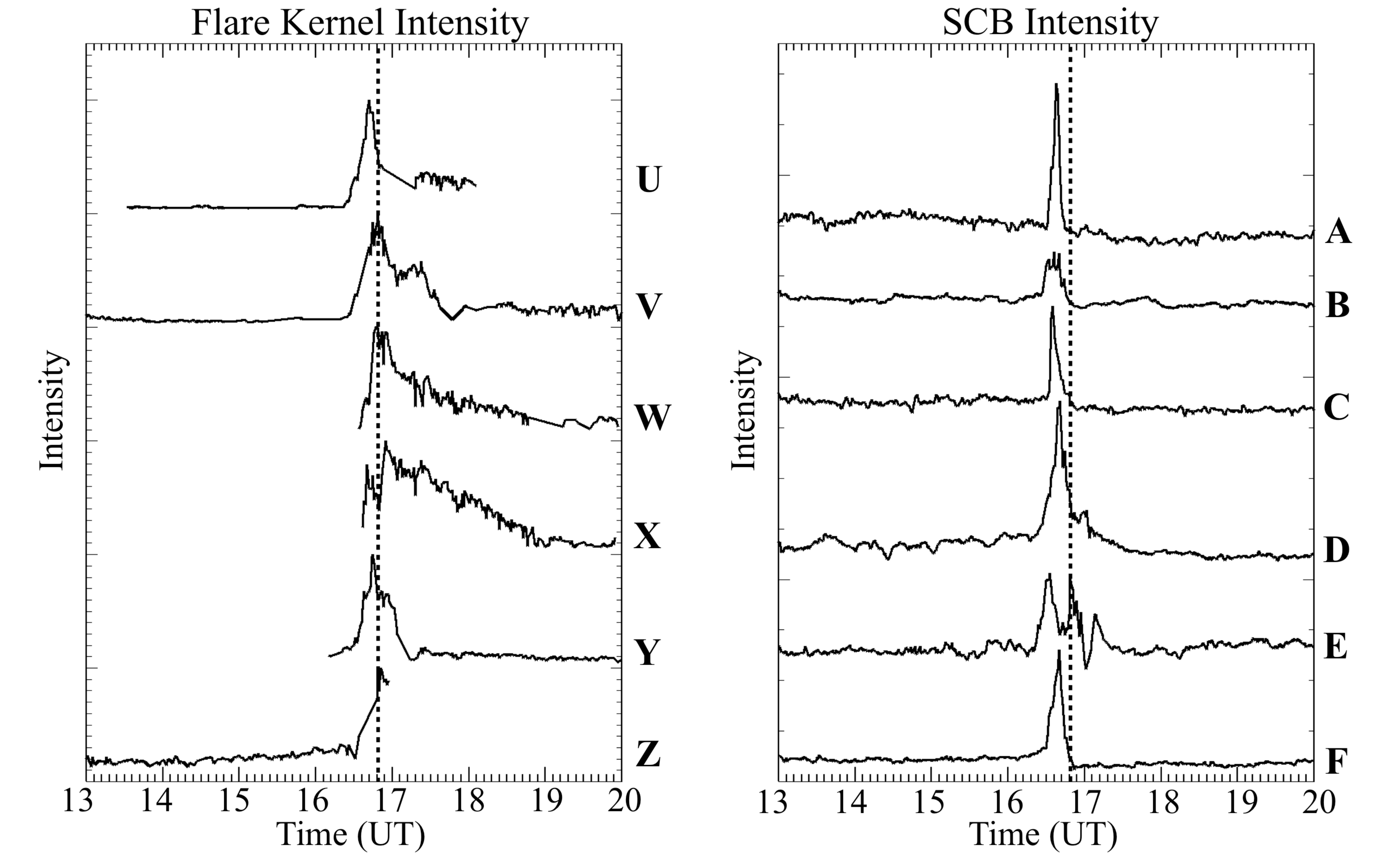}
              }
              \caption{H$\alpha$ normalized intensity curves for each of the kernels indicated in Figure~\ref{Kernel_map} by the associated letters. On the left, the tracked flare kernels are shown.  The normalized intensity curves for the SCB kernels, shown on the right, are stationary and therefore tracked throughout the flaring window. In both panels, the vertical dashed line indicates the time at which the flare intensity peaks (16:49 UT) derived from integrating all flare kernel intensities at each time step.
                             }
   \label{Kernel_curves}
   \end{figure}

SCBs, although related to erupting flare ribbons, are distinctly different from the flare kernels discussed in Section~\ref{S-Flare}. Six SCBs are chosen from the 13 May 2005 event as an example of these ephemeral phenomena. The locations of these six events are letters A - F in Figure~\ref{Kernel_map}.  The \Ha normalized intensity of each of these kernels is shown in Figure~\ref{Kernel_curves}. The SCB intensity curve is significantly different from the flare kernel curves shown on the left side of the figure. SCB curves are impulsive; they have a sharp peak and then return to background intensity in the span of about 12 minutes. Nearly all of the SCBs shown here peak in intensity before the peak of the flare intensity curve, shown as a vertical dashed line in Figure~\ref{Kernel_curves}. SCBs B and E both appear to have more internal structure than the other SCBs and last noticeably longer. This is most likely caused by several unresolved SCB events occurring in succession. 

\begin{figure}    
   \centerline{\includegraphics[width=1.0\textwidth,clip=,angle=0]{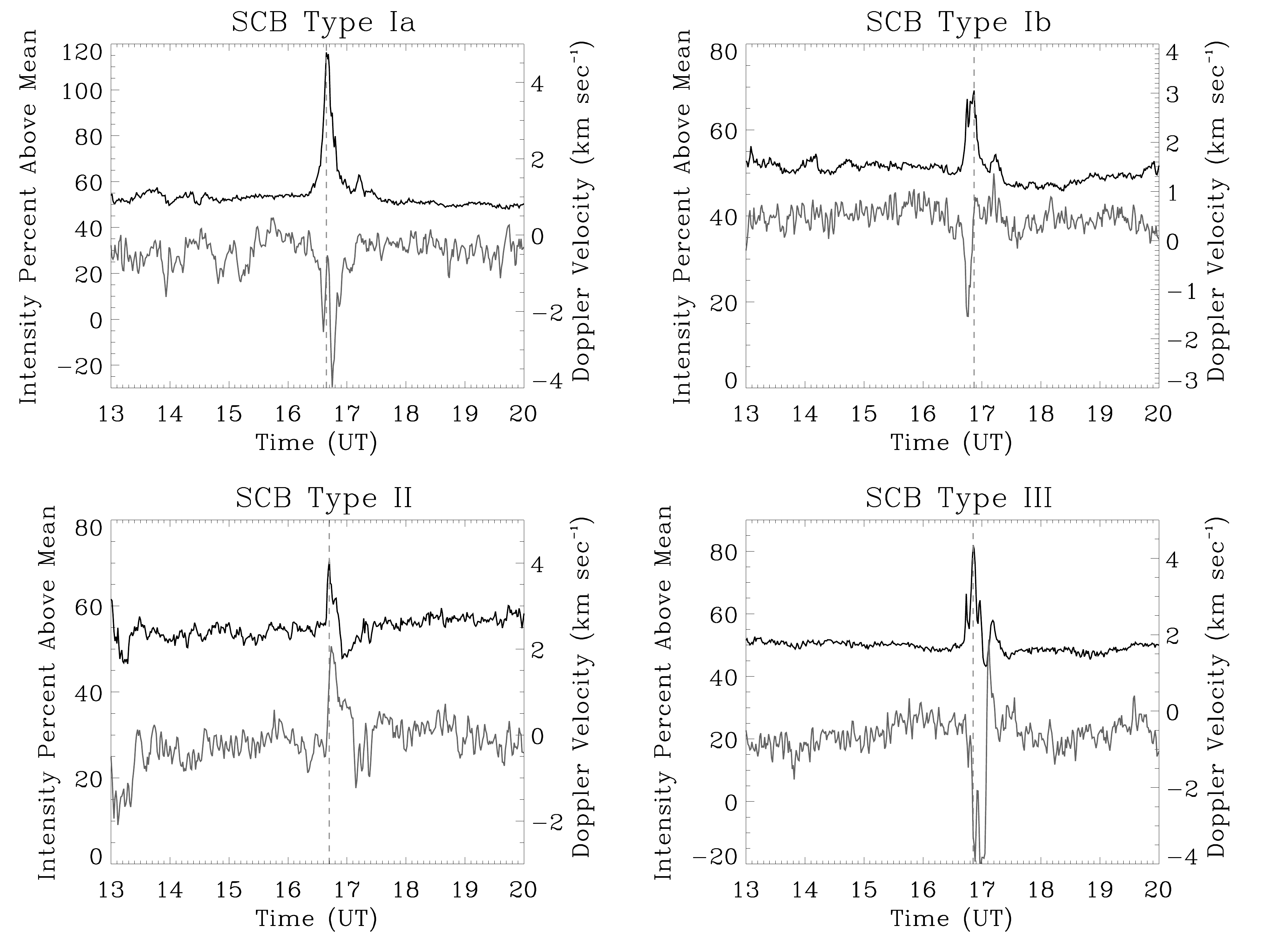}
              }
              \caption{The line center intensity and Doppler velocity measurements for three different types of SCBs observed in both 2005 flares. The vertical dashed line marks the peak H$\alpha$ intensity of the associated SCB. A negative Doppler velocity is away from the observer and into the Sun. 
                             }
   \label{SCB_types}
   \end{figure}

Nearly all SCBs have \Ha line center intensities 40 -- 60\% above the mean intensity of the quiet chromosphere (Figure~\ref{SCB_types}).  From this increased initial intensity, SCBs brighten to intensities 75 -- 130\% above the quiet Sun. 

Examining Doppler velocity measurements from the SCB locations reveals three distinct types of SCBs (Figure~\ref{SCB_types}). A type Ia SCB has an impulsive intensity profile and an impulsive negative Doppler profile that occurs simultaneously or a few minutes after the peak brightening. In this study, a negative velocity is associated with motion away from the observer and into the Sun. A type Ib SCB has a similar intensity and Doppler profile as a type Ia but the timing of the impulsive negative velocity occurs several minutes before the peak intensity of the SCB. The type Ib SCB shown in Figure~\ref{SCB_types} has a negative velocity that peaks 10 minutes before the peak intensity and returns to a stationary state before the \Ha line center intensity has decayed. A type II SCB has a positive Doppler shift perturbation that often lasts longer than the emission in the \Ha intensity profile. The timing of both are nearly coincidental. A type III SCB demonstrates variable dynamics. It has a broad \Ha intensity line center with significant substructure.  A type III SCB begins with a negative Doppler profile much like a type I. Before the negative velocity perturbation can decay back to continuum levels, there is a dramatic positive velocity shift within three minutes with an associated line center brightening.  In all types of SCBs the typical magnitude of Doppler velocity perturbation is between 2 and 5 \kms in either direction perpendicular to the solar surface.

Of all off-flare compact brightenings detected using the automated techniques, 59\% of SCBs have no discernible Doppler velocity in the 6 May event and 21\% in the 13 May event.  Out of the SCBs that do have an associated Doppler velocity, the 6 May event has 35\% of type I, 54\% of type II, and 11\% of type III. The 13 May event has 52\% of type I, 41\% of type II, and 17\% of type III. When totaling all of the Doppler velocity associated SCBs between both events, 41\% are of type I, 46\% of type II, and 13\% of type III. Both line center brightenings and Doppler velocities are filtered such that positive detections have at least a two standard deviation peak above the background noise determined in the candidate detection \citep{Kirk2011}. 

\subsection{SCBs in aggregate} 
\label{S-totalSCB}

\begin{figure}    
   \centerline{\includegraphics[width=0.70\textwidth,clip=,angle=0]{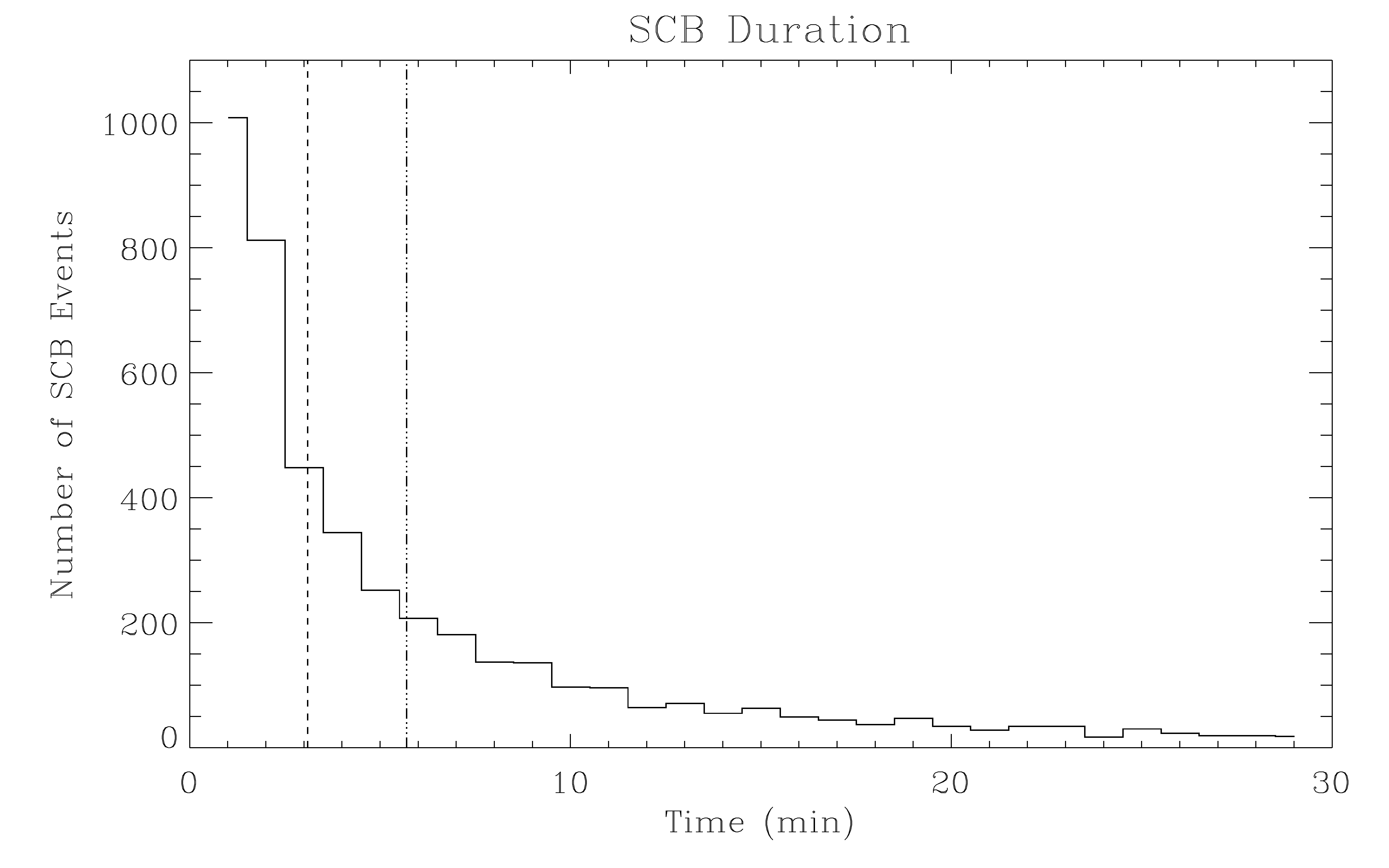}
              }
              \caption{A histogram of the duration of all SCB detections in all three events. The SCBs have a measured mean duration of 5.7 minutes (dot-dashed line) and a median duration of 3.1 minutes (dashed line).                    
                         }
   \label{SCB_duration}
   \end{figure}

Examining SCBs as a total population, the intensity brightenings have a median duration of 3.1 minutes and a mean duration is 5.7 minutes (Figure~\ref{SCB_duration}). The duration is characterized by the full-width half-maximum of the SCB intensity curve (examples of these curves are shown in Figure~\ref{Kernel_curves}).  A histogram of the distribution of the number of SCB events as a function of duration shows an exponential decline in the number of SCB events between 2 and 30 minutes (Figure~\ref{SCB_duration}).  The duration of SCBs is uncorrelated with both distance from flare center and the peak intensity of the SCB.

\begin{figure}    
   \centerline{\includegraphics[width=0.70\textwidth,clip=,angle=0]{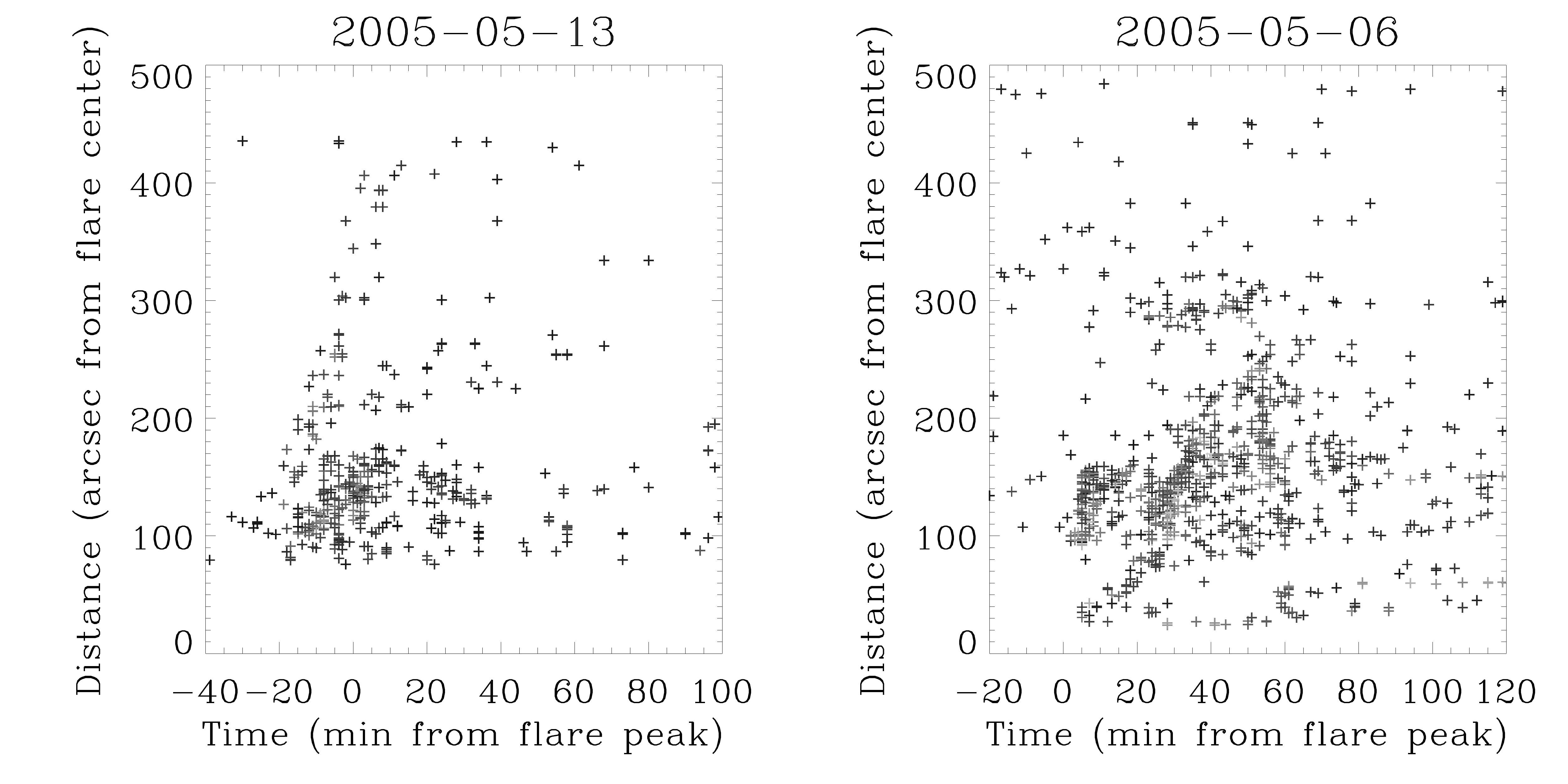}}
   \centerline{\includegraphics[width=0.70\textwidth,clip=,angle=0]{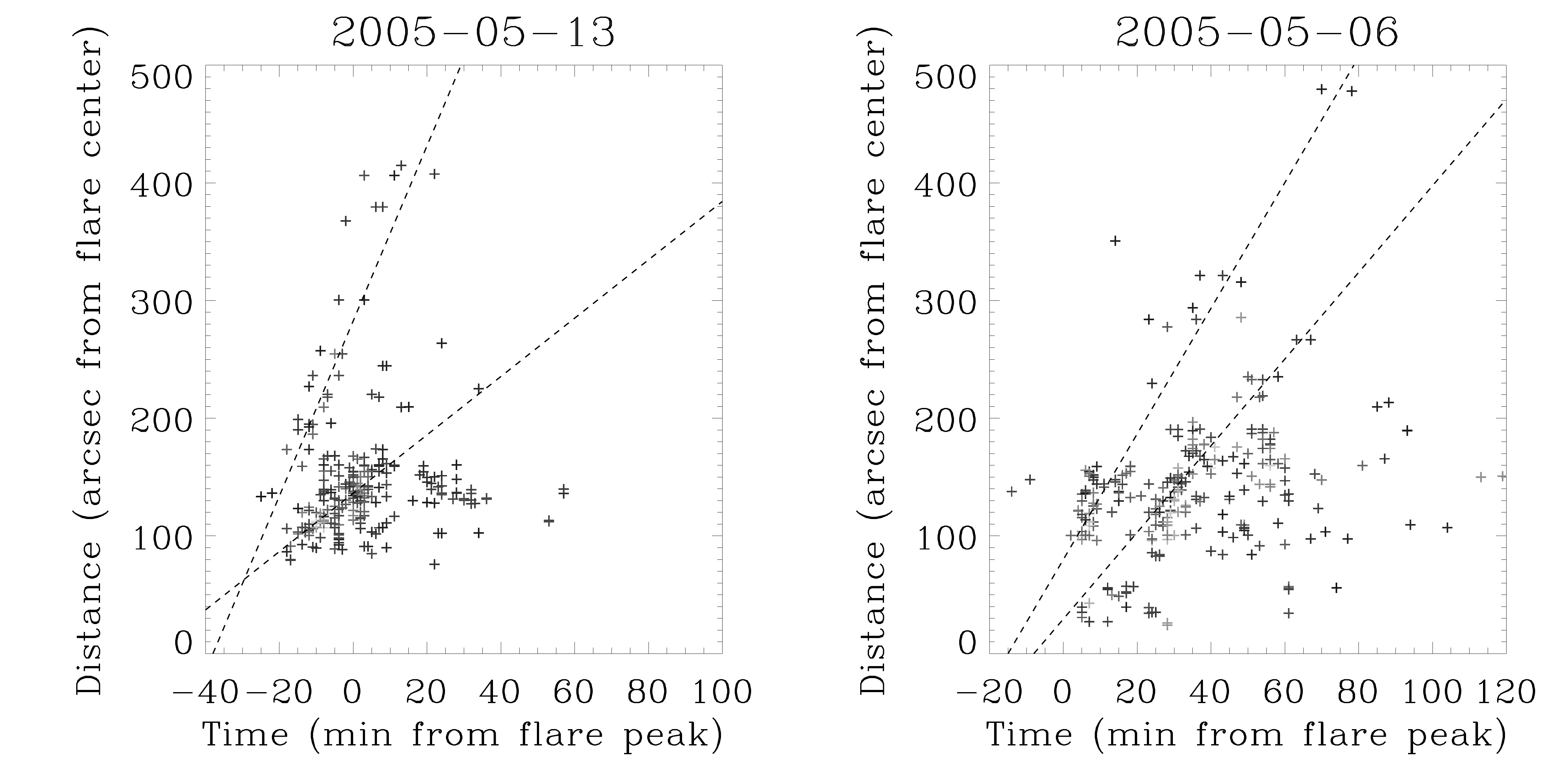}}   
              \caption{Derived statistics for all SCBs for the 6 May 2005 and 13 May 2005 events. All panels show the distance at which the SCB occurs versus time from the H$\alpha$ flare peak. The color of each plotted point is representative of its relative intensity; the lighter points are higher intensity detections. Bottom: the same plots as the top two panels except with a Doppler filter applied -- each point had to have a SCB Type I Doppler shift associated with the line center brightening.  The dashed lines are fit to the two visually distinct groups in each event. 
	}
   \label{SCB_stats}
   \end{figure}

Examining individual SCBs' distance from the flare center as a function of time provides a way to extract the propagation trends of SCBs around the flare (Figure~\ref{SCB_stats}). The bulk SCBs appear between 1.2 and 2.5 Mm away from flare center in a 1 hour time window. At the extremes, SCBs are observed at distances up to $\approx 5$ Mm and several hours after the flare intensity maximum.  On the top two panels of Figure~\ref{SCB_stats}, the shade of the mark corresponds to the intensity of the SCB measured where the lighter the mark, the brighter the SCB. The center of the flare is determined by first retaining pixels in the normalized dataset above 1.35. Then all of the images are co-added and the center of mass of that co-added image is set to be the ``flare center." The SCBs tend to clump together in the time-distance plot in both the 6 May and 13 May events. Generally, the brighter SCBs are physically closer to the flare and temporally occur closer to the flare peak. This intensity correlation is weak and qualitatively related to distance rather than time of brightening. Statistics from the November 9, 2004 event are dominated by noise. 

Identifying SCBs with negative Doppler shifts within 3 minutes of the \Ha peak intensity limits SCBs to type I only.   With this method of filtering, some trends become more apparent in a time-distance plot of SCBs (Figure~\ref{SCB_stats}, bottom panels).  In this view of SCBs, they tend to cluster together in time as well as distance, with two larger groups dominating the plots. 

To fit a slope to these populations, a forward-fitting technique is employed similar to a linear discriminate analysis. This method requires the user to identify the number of groups to be fitted and the location of each group. The fitting routine then searches all linear combinations of features for the next ``best point" to minimize the chi-square to a regression fit of the candidate group. Repeating this method over all the points in the set produces an ordered set of points that when fit, have an increasing chi-square value. A threshold is taken where the derivative of the chi-square curve increases to beyond one standard deviation, which has the effect of identifying where the chi-square begins to increase dramatically. Running this routine several times minimizes the effect of the user and provides an estaminet of the error associated with the fitted line. This method has two caveats. First, this fitting method relies on the detections having Poissonian noise. This is not necessarily correct since the detection process of compact brightenings introduces a selection bias. Second, the fitting method makes the assumption that no acceleration occurs in the propagation of SCBs. This is a reasonable approximation but from studies of Moreton waves in the chromosphere \citep[e.g.][]{Bala2010}, a constant velocity propagation is unlikely. 

Applying the forward-fitting technique these two data sets yields two propagation speeds: a fast and a slow group. The 6 May event has propagation speeds of  $71 \pm 2$ \kms and $49 \pm 2$ \kms. The 13 May event has propagation speeds of $106 \pm 7$ \kms and $33 \pm 4$ km s$^{-1}$. Beyond these detections, the 13 May event has a third ambiguous detection with a propagation speed of  $-7 \pm 5$ km s$^{-1}$. The implications of these statistics are discussed in Section~\ref{S-Discussion}. 

\begin{figure}    
   \centerline{\includegraphics[width=1.0\textwidth,clip=,angle=0]{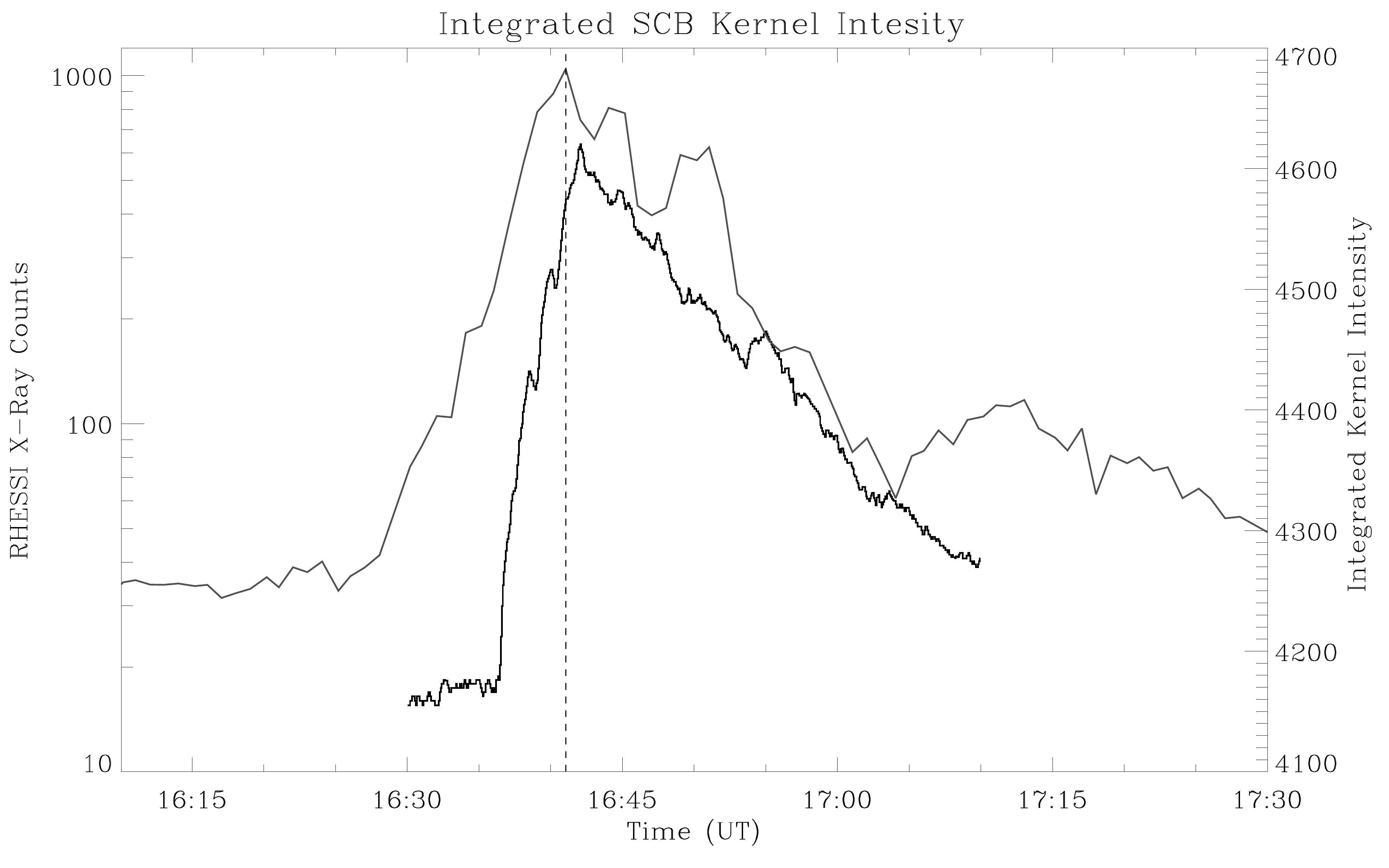}
              }
              \caption{A plot showing the RHESSI non-thermal 25 -- 50 keV intensity in black and the H$\alpha$ SCB integrated intensity curve in gray for the 13 May 2005 flare. The vertical dashed line indicates the peak of the integrated SCB curve. 
                                           }
   \label{SCB_RHESSI}
   \end{figure}

The x-ray intensity measured with RHESSI temporally covers a part of the flaring event on 13 May 2005. RHESSI data coverage begins at 16:37 UT during the impulsive phase of the flare. Comparing the integrated x-ray intensity curve between 25 -- 50 keV to the aggregate of SCB intensities integrated over each minute yields some similarities (Figure~\ref{SCB_RHESSI}). The x-ray intensity peaks about a minute after the integrated SCBs reach their maximum intensity. The decay of both the x-ray intensity and SCB integrated curve occur approximately on the same time-scale of $\approx 50$ minutes. 

\section{Physical interpretation of results}
\label{S-Interp}
From the results of the tracking method, we can gain insight into the physical processes associated with SCBs and flares. First, the flare ribbons studied are both spatially and temporally clumped into discrete kernels. The relatively smooth ribbon motions observed in \Ha and the nearly ideal exponential decay characterized in x-ray flux measurements belie the substructure that makes up a two-ribbon flare. This is discussed in Section~\ref{S-Model_flare}. Second, the brightening in the flare ribbons is caused by a distinctly different physical process than the one causing SCBs. In this sense, SCBs are different than micro-flares. Our interpretation of the physical origin of SCBs is addressed in Section~\ref{S-Model_SCB}. 

\subsection{Two ribbon chromospheric flares} 
\label{S-Model_flare}
Flare kernels are observed to appear and disappear as the underlying flare ribbons evolve, as discussed in Section~\ref{S-Individual}. Examining individual kernel structure suggests there is substructure within a flare ribbon whose elements impulsively brighten and dim within the brightness that encompasses the intensity curve. These results support the premise that flares are made up of several magnetic field lines reconnecting~\citep[e.g.,][]{Priest2002}.  There is no evidence to claim an individual flare kernel is directly tracking a loop footpoint. Within a tracked flare kernel, multiple coronal reconnection events are superimposed to produce the observed asymmetries a single flare kernel's light curve. 

\citet{Maurya2010} tracked subsections of the 28 October 2003 X17 flare as it evolved. They reported peak speeds ranging from $\sim 10 - 60$ km s$^{-1}$ (depending on the spatially tracked part of the flare) over the observed span of 13 minutes.  \citet{Maurya2010} also reported that the total apparent distances the ribbons traveled were $\sim 10^4$ km. The 28 October flare is several orders of magnitude greater in GOES x-ray intensity than the flares considered for this study. In the present study, despite this difference, peak speeds of flare kernels observed in this work are measured at $\sim 2.3$ km s$^{-1}$ and the mean velocity of all flare kernels is $\sim 0.2$ km s$^{-1}$. The maximum distance flare kernels traversed was $\sim 2 \times 10^4$ km and the average distance traveled was $\sim 5 \times 10^3$ km; both of these values are similar to the 28 October 2003 event. Although the velocities of the flare ribbons studied here are at least an order of magnitude less, the two to three orders of magnitude difference in the GOES peak intensity between the 28 October flare and the flares studied here implies that the apparent velocities and distance traveled by flare ribbons do not scale linearly with the strength of the eruption.

The bulk of the apparent motion and acceleration in the flare ribbons are observed after the peak of the flare intensity. Integrating this into the dynamical model of a two-ribbon flare~\citep{Demoulin1988} implies that the peak intensity of the flare occurs in the low lying arcade and loses intensity as the x-point reconnection progresses vertically to higher levels~\citep{Baker2009}. The measured divergence of the ribbons dominate the motion but there is significant apparent flow tangential to the flare ribbons. This predicates that there is lateral propagation to the x-point as well as vertical propagation. 

\subsection{A physical model of SCBs}
\label{S-Model_SCB}

\begin{figure}    
   \centerline{\includegraphics[width=1.0\textwidth,clip=,angle=0]{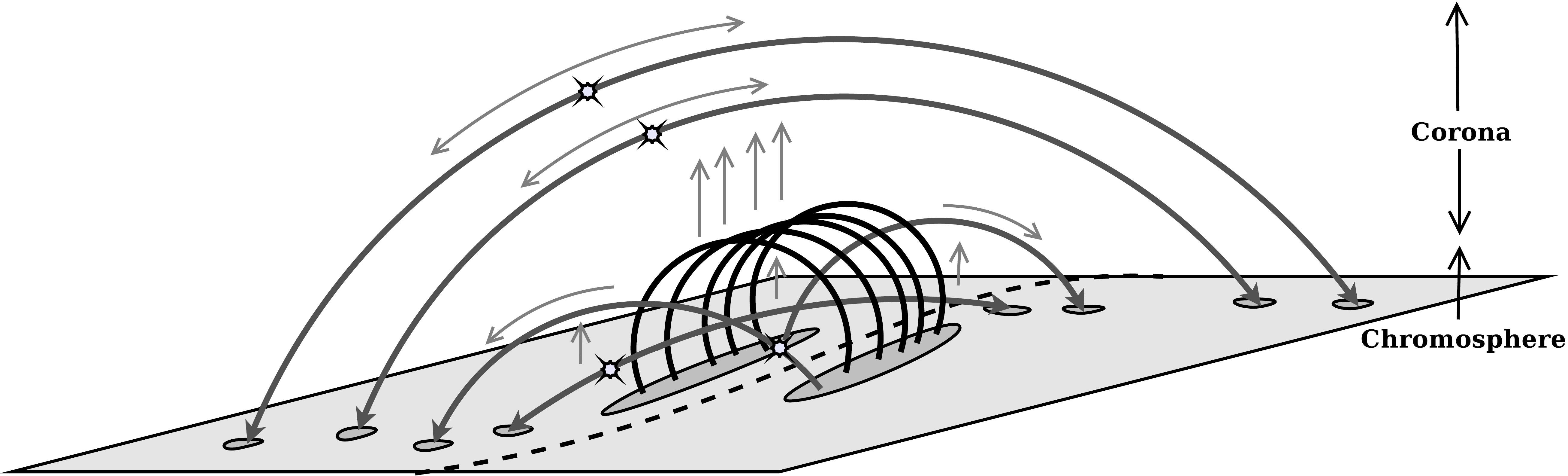}
              }
              \caption{A cartoon model of the physical cause of SCBs. A dashed line marks the neutral line in this model. The central loops represent the emerging two ribbon flare arcade. The arches represent field lines connecting outside of the flare ribbons. The arrows on top of both the loops and arches show the direction trapped plasma flows as the flare begins to erupt and the stars suggest the location where the plasma is disrupted. Once perturbed, the trapped plasma streams down the loop lines and impacts the chromosphere causing the H$\alpha$ intensity brightening.  The different orientations of off-flare loops accounts for the different propagation speeds of SCBs observed. 
                     }
   \label{SCB_loops}
   \end{figure}

Figure~\ref{SCB_loops} is a proposed phenomenological model of the overlying physical topology. SCBs are hypothesized to be caused by electron beam heating confined by magnetic loop lines over-arching flare ribbons. These over-arching loops are analogous to the higher lying, unsheared tethers in the breakout model of CMEs \citep{Antiochos1999}.  As the flare erupts, magnetic reconnection begins in a coronal x-point. The CME escapes into interplanetary space, the remaining loop arcade produces a two ribbon flare, and the tethers reconnect to a new equilibrium position.  The tether reconnection accelerates trapped plasma which impacts the denser chromosphere causing observed brightening. This description implies the driver of SCBs is the eruption of an CME. \citet{Bala2006} found a strong correlation between CMEs and the presence of SCBs. 

A simple loop configuration without localized diffusion or anomalous resistance implies that the length of the loop directly proportional to the travel time of the electron beam. This means that the different propagation groups observed in Figure~\ref{SCB_stats} result from different physical orientations of over-arching loops. The time it takes to observe a chromospheric brightening can therefore be described as:
\begin{equation}
\label{SCB_formula}
\tau_{\rm SCB}=\gamma \frac{L}{v_e},
\end{equation}
where $\tau_{\rm SCB}$ is the time it takes for accelerated plasma to travel from the coronal x-point, along the magnetic loop lines, impact the chromosphere, and produce a brightening, $L$ is the length of the tether before reconnection, $v_e$ is the electron velocity along the loop line, and $\gamma$ is a function of changing diffusion and resistance between chromosphere and corona. In physical scenarios, $v_e$ is likely a combination of the Alfv\'en speed and the electron thermal velocity. 

The slopes plotted in Figure~\ref{SCB_stats} show two different propagation speeds for SCBs: $33 - 49$ km s$^{-1}$ and $71 - 106$ km s$^{-1}$. The sound speed in the chromosphere is approximately $c_s\approx10$ km s$^{-1}$ \citep{Nagashima2009}, while the Alfv\'en speed in the upper chromosphere is approximated to be between $v_A \approx 10 - 100$ km $s^{-1}$  \citep{Aschwanden2005}. Both of the SCB propagation speeds are significantly above the approximated sound speed, however they fall reasonably well into the range of the Alfv\'en speed. Assuming the propagation speed of non-thermal plasma along the loop line is consistent between flare loops, the different propagation speeds would then imply different populations of loops being heated as the flare erupts.   

Examining single SCBs, the coincident Doppler recoil with the \Ha intensity presents a contradiction. If an SCB is an example of compact chromospheric evaporation \citep[e.g.][]{Dennis1989}, then the only Doppler motion should be outward, the opposite of what is observed. Figure~\ref{SCB_cartoon} presents a possible solution to this.  Since the scattering length of electrons is significantly smaller than that of protons, the electron beam impacts the mid chromosphere and deposit energy into the surrounding plasma while the protons penetrate deeper. This deposited heat cannot dissipate effectively through conduction or radiation and thus expands upward into the flux tube. To achieve an expansion with rates of a few km s$^{-1}$ as observed, the chromospheric heating rate must be below $E_H \le 10^{10}$ erg cm$^{-2}$ s$^{-1}$ \citep{Fisher1984}.  As a reaction to this expansion, a reaction wave propagates toward the solar surface in the opposite direction of the ejected plasma. Since the Doppler measurements are made in the wings of the \Ha line, the location of the Doppler measurement is physically closer to the photosphere than the \Ha line center. Thus the observer sees the heating of the \Ha line center and coincidentally observes the recoil in the lower chromosphere. 

In type II SCBs an up-flow is observed. This is an example of a classic chromospheric evaporation where the heated plasma in the bulk of the chromosphere is heated and ablated back up the flux tube.  In contrast, the type III class of SCB shown in Figure~\ref{SCB_types} present an interesting anomaly to both the other observed SCBs and the model proposed in Figure~\ref{SCB_cartoon}. Since there is an initial down-flow, the beginning state of type III SCBs are similar to type Is. As the recoil is propagated, a continual bombardment of excited plasma (both protons and electrons) impacts the lower chromosphere, causing ablation and changing the direction of flow.

\section{Summary and Conclusions} 
\label{S-Discussion}

Two conclusions can be made about the flares studied in this work: first, the asymmetrical motions of the flare ribbons imply the peak flare energy occurs in the low lying arcade, and second, flare related SCBs appear at distances on the order of $10^5$ km and have properties of sites of compact chromospheric evaporation. \citet{Maurya2010} estimated reconnection rates from a measurement of the photospheric magnetic field and the apparent velocity of chromospheric flare ribbons. Therefore, it would be possible to estimate the reconnection rates using this technique with the addition of photospheric magnetograms. Associating vector magnetograms with this technique and a careful consideration of the observed Doppler motions underneath the ribbons would provide a full 3D method for estimating the Lorentz force for subsections of a flare ribbon.   

\begin{figure}    
   \centerline{\includegraphics[width=0.70\textwidth,clip=,angle=0]{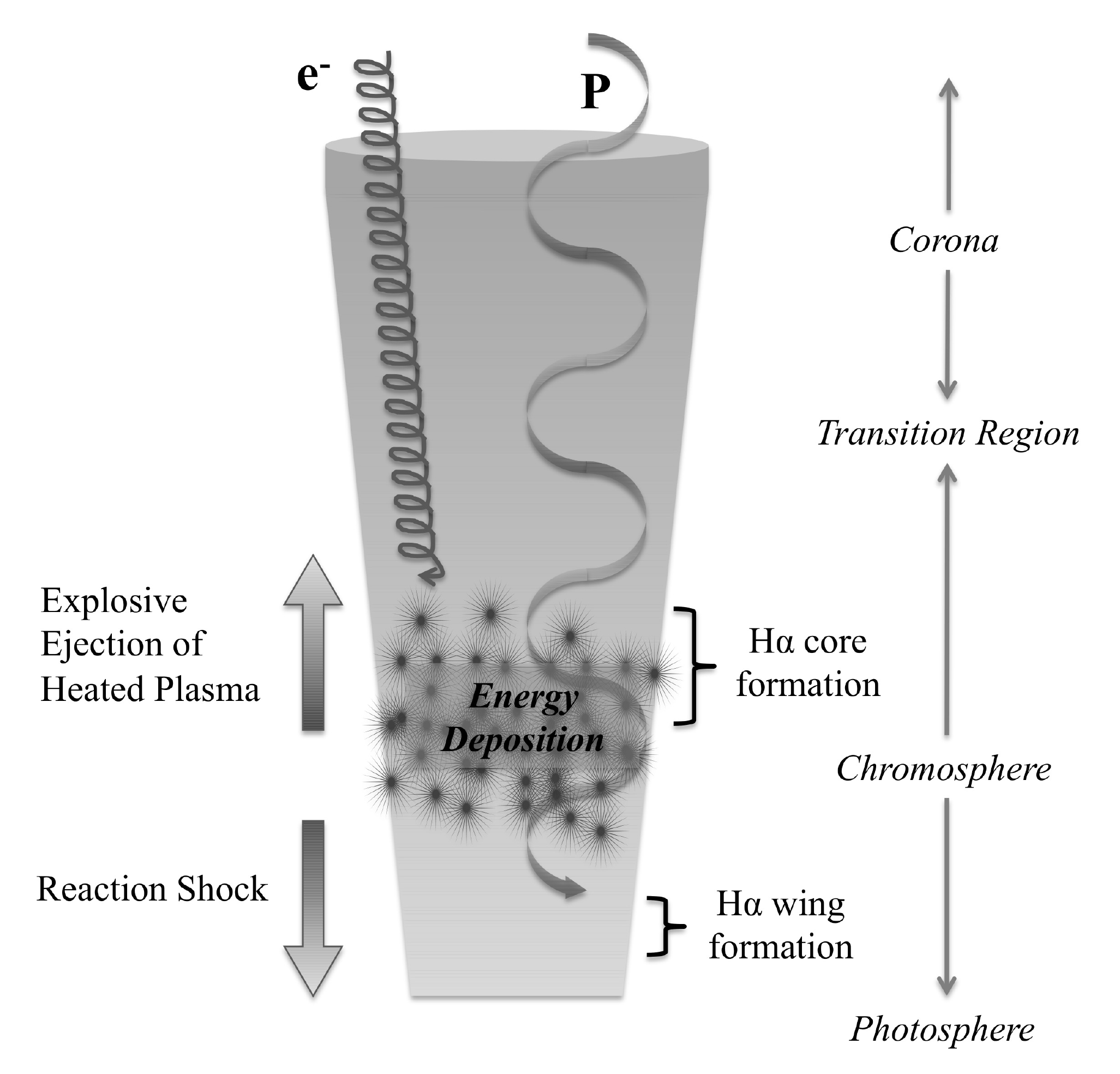}
              }
              \caption{A diagram of the physical dynamics occurring in a single SCB. Electrons and protons accelerated by magnetic reconnection further up in the corona come streaming down the flare loop. Since the mean free path of electrons is significantly less, the electrons deposit energy into the upper chromosphere. Unable to radiate the energy as heat, the chromosphere explosively responds by sending material back up the flux tube (chromospheric evaporation) and a reaction shock propagates towards the photosphere. This reaction shock explains the velocities observed in Type I SCBs. 
                             }
   \label{SCB_cartoon}
   \end{figure}

SCBs originate during the impulsive rise phase of the flare and often precede the \Ha flare peak. SCBs are found to appear with similar qualities as compact chromospheric ablation confirming the results of \citet{Pevtsov2007}. The heuristic model proposed in Figure~\ref{SCB_cartoon} requires unipolar magnetic field underneath SCBs. The integration of high resolution magnetograms and a field extrapolation into the chromosphere would confirm this chromospheric evaporation model. One consequence of chromospheric evaporation is that the brightening should also be visible in EUV and x-ray observations due to non-thermal protons interacting with the lower chromosphere or photosphere. 

SCBs are a special case of chromospheric compact brightening that occur in conjunction with flares. The distinct nature of SCBs arises from their impulsive brightenings, unique Doppler velocity profiles, and origin in the impulsive phase of flare eruption. These facts combined demonstrate that SCBs have a non-localized area of influence and are indicative of the conditions of the entire flaring region. They can possibly be understood by a mechanism in which a destabilized overlying magnetic arcade accelerates electrons along magnetic tubes that impact a denser chromosphere to result in an SCB. This distinguishes SCBs from the flare with which they are associated.  In a future work, we plan to address physical mechanisms to explain the energetic differences between SCBs and flares to explain the coupled phenomena. 

\acknowledgments
The authors thank: (1) USAF/AFRL Grant FA9453-11-1-0259, (2) NSO/AURA for the use of their Sunspot, NM facilities, (3) AFRL/RVBXS, and (4) Crocker and Weeks for making their algorithm available online.

\bibliography{SCB_Methods_Bibliography.bib}

\clearpage

   \clearpage
   
   \clearpage
   \clearpage
    \clearpage
 
      \clearpage
      \clearpage
      
      \clearpage

   \clearpage
   \clearpage
 
   \clearpage
   
       \clearpage 

       \clearpage 
 
       \clearpage 

 
       \clearpage

       \clearpage 
       \clearpage 

       \clearpage 
       
\end{document}